\documentclass[11pt]{article}
\usepackage{afterpage,amsmath,amssymb,epsf,psfig,graphicx}
\usepackage{psfrag}

\newcommand{\beq}{\begin{equation}}
\newcommand{\eeq}{\end{equation}\noindent}
\newcommand{\bear}{\begin{eqnarray}}
\newcommand{\eear}{\end{eqnarray}\noindent}

\newenvironment{mydescription}[1]
 {\begin{list}{}{
  
  \setlength{\leftmargin}{10pt}
  \setlength{\itemindent}{10pt}
  \setlength{\rightmargin}{10pt}}}
  {\end{list}}

\begin{document}

\begin{flushright}
BI/TH-98/38 \\
December 1998
\end{flushright}

\begin{center}

\vspace{24pt}
{\LARGE \bf The Weak-Coupling Limit of\\[5pt]
            \hspace{5pt}Simplicial Quantum Gravity} 
\vspace{24pt}

{\Large \sl G.~Thorleifsson, P.~Bialas\footnote{Permanent 
 address: Institute of Comp.~Science Jagellonian University, 
 30-072 Krakow, Poland} \,{\rm and}\, B.~Petersson} \\
\vspace{10pt}

Facult\"{a}t f\"{u}r Physik, Universit\"{a}t Bielefeld
D-33615, Bielefeld, Germany  \\
\vspace{10pt}

\begin{abstract}
In the weak-coupling limit, $\kappa_0 \rightarrow \infty$, 
the partition function of simplicial quantum gravity
is dominated by an ensemble of triangulations with the 
ratio $N_0/N_D$ close to the  upper kinematic limit. 
For a combinatorial triangulation of the $D$--sphere this 
limit is $1/D$.  Defining an ensemble of {\em maximal} 
triangulations, i.e.\ triangulations that have the maximal possible 
number of vertices for a given volume,
we investigate the properties of this ensemble in three dimensions 
using both Monte Carlo simulations and a strong-coupling 
expansion of the partition function, both for pure simplicial
gravity and a with a suitable modified measure.  For the latter
we observe a continuous phase transition to a {\it crinkled} phase
and we investigate the fractal properties of this phase.
\end{abstract}

\end{center}
\vspace{15pt}

\section{Introduction}

Discretized models of $D$--dimensional Euclidean
quantum gravity, known as simplicial gravity or 
dynamical triangulations, have been extensively studied
in the past decade. The two dimensional model was proposed
quite some time ago in Ref.~\cite{dt2d}, and generalized to
three and four dimensions in Ref.~\cite{dt34d,1st3D}. 
The model is defined by the grand-canonical 
partition function
\begin{eqnarray}	
{\cal Z}(\mu,\kappa_0)=\sum_{N_D} e^{\textstyle \;-\mu N_D}Z(\kappa_0,N_D)
\label{part.grand}
\end{eqnarray}
where $\mu$ and $\kappa_0$ are the discrete cosmological 
and inverse Newton's constants, $N_D$ is the number of
$D$--dimensional simplices, and $Z(\kappa_0,N_D)$ 
is the canonical partition function
\begin{equation}
  Z(\kappa_0,N_D) \;=\; 
   \sum_{N_0} \,{\rm e}^{\textstyle \;\kappa_0 N_0} \; W_D(N_0,N_D)\;.
 \label{part.canon}
\end{equation}
The micro-canonical partition function, $W_D$, is defined as
\begin{equation}
  W_D(N_0,N_D) 
  \;= \!\!\sum_{T \in {\cal T}(N_0,N_D)} \; \frac{1}{C_T}  \, 
   \,=\, \frac{1}{N_0!}  \, \sum_{L \in {\cal L}(N_0,N_D)} 1\;,
  \label{part.micro}
\end{equation}
where the sum is over an ensemble of combinatorial 
triangulations ${\cal T}$ with a spherical topology and
with $N_0$ vertices and $N_D$ $D$--simplices. 
For $D > 2$ these two numbers are independent.
The symmetry factor $C_T$ of a triangulation $T$ is given
by the number of equivalent labeling of the vertices. 
Including this factor is equivalent to summing over all 
{\em labeled} triangulations ${\cal L}(N_0,N_D)$. 

As a statistical system simplicial gravity
displays a wealth of intriguing features. In $D > 2$ it
exhibits a geometric phase transition 
separating a strong-coupling {\it crumpled} phase, where the 
internal geometry of the triangulations collapses, 
from a weak-coupling {\it elongated} phase characterized by 
triangulations consisting of ``bubbles'' glued together into a 
tree-like structure or branched polymers \cite{lat98}. 
Both in three \cite{1st3D} and four dimensions \cite{1st4D}
this transition is discontinuous.

In the absence of a continuous phase transition with a divergent 
correlation length any sensible continuum limit where a
theory of gravity might be defined appears to be ruled out.  
This has led to the studies of modified models of simplicial 
gravity, e.g.\ by changing the relative weight of 
the triangulations with a modified measure \cite{enso}, 
or by coupling matter fields to the theory \cite{matter}.
Suitable modified the phase structure of the model changes;  
the polymerization of the geometry in the weak-coupling limit 
--- the elongated phase --- is suppressed 
and a new {\it crinkled} phase appears \cite{us4d}.
The crinkled phase is separated from the other two phases
by lines of either a soft continuous phase transitions,
or possible a cross-over (a divergent specific heat
is not observed).  Qualitatively the same phase structure,  
shown schematically in Figure~1, is observed both in three and four 
dimensions.

Both in the elongated and in the crinkled phase
the canonical partition function is seen to have the 
asymptotic behavior
\begin{equation}
 \label{def.gamma}
 Z(\kappa_0,N_D) \;\sim\; {\rm e}^{\textstyle \; \mu_c N_D}\;
  N_D^{\textstyle \gamma-3}\;,
\end{equation}
which defines the string susceptibility exponent $\gamma$.
In the crumpled phase, on the other hand, the sub-leading
behavior is exponential, corresponding to $\gamma = -\infty$.

\begin{figure}
 \epsfxsize=4in \centerline{ \epsfbox{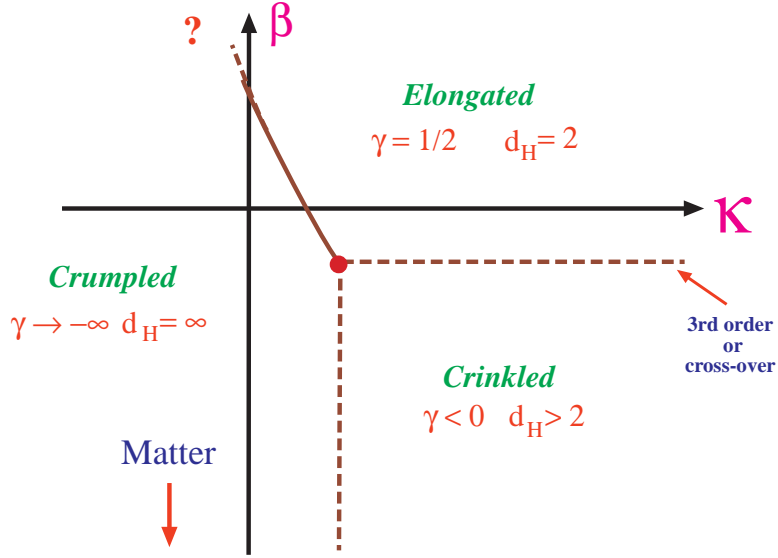}}
  \caption[fig.phase]{\small A schematic phase diagram of modified
    simplicial quantum gravity for $D> 2$.  The solid line
    indicates  discontinuous phase transitions,
    the dashed lines are either soft continuous transitions
    or a cross-over.}
  \label{fig.phase}
\end{figure}

Many of the properties of the model Eq.~(\ref{part.grand}), in
the weak-coupling regime $\kappa_0~\!>~\!\kappa_0^c$, are
reflected by triangulations close to the upper kinematic bound
$N_0/N_D = n_0^{max}$.  This is evident both from Monte Carlo
(MC) simulations and from the analysis of a strong-coupling expansion
(SCE) of the partition function.  
For combinatorial triangulations, where each simplex is
uniquely defined by a set of $\bigl(D+1\bigr)$ distinct vertices, 
the upper kinematic bound is $n_0^{\it max} = 1/D$. 
Although this bound is only saturated in the thermodynamic limit,
$N_D \rightarrow \infty$, for each finite volume there exist
triangulations with a well-defined maximal number
of vertexes $N_0^{\it max}(N_D)$.  The ensemble of these triangulations,
the {\it maximal} ensemble ${\cal T}_{\it max}$, 
defines the micro-canonical partition function
\begin{equation}
 \label{part.max}	
  W_{max}(N_D) \;\equiv \; W_D \bigl (N_0^{max}(N_D),N_D \bigr )
  \;=\; \sum_{T \in {\cal T}_{max}(N_D)} \;\frac{1}{C_T}\;.
\end{equation}

In this paper we investigate the properties of
the maximal ensemble, mostly restricted to three dimensions,  
using both MC simulations and a SCE, i.e.\ a direct enumeration 
of the micro-canonical partition function Eq.~(\ref{part.max}).
This investigation is separated into two parts. 
In the first part we study the pure (un-modified)
ensemble and show that $W_{max}(N_3)$ effectively separates
into {\it three} distinct series (or two series in $4D$)
as in general there are several different volumes $N_D$ 
corresponding to a given $N_0^{max}$.
One of those series is an ensemble of {\it stacked spheres},
a particular simple class of triangulations which
can be enumerated exactly.  This we do in Section~3.

The other two series can be considered as ``almost'' stacked 
spheres, i.e.\ with one, or two, {\it defects} respectively.  
Heuristic arguments, supported by numerical evidence, 
show that although the leading asymptotic
behavior of the three series is identical, the effect of 
inserting a defect increases the exponent $\gamma$
by $+1$ (or $+2$) compared to stacked spheres.
This is discussed in Section~4.

In the second part of the paper, Section~5, we study the model 
Eq.~(\ref{part.max}) with a modified measure,
\begin{equation}
 \label{part.measure}
 W_{max}(N_D,\beta) \;=\; 
 \sum_{T \in {\cal T}_{ max}} \;\frac{1}{C_T} \;
 \prod_{i=1}^{N_0} \;q_i^{\;\beta} \;.
\end{equation}
Here $q_i$ is the {\it order} of a vertex $i$,
i.e.\ the number of simplices incident to that vertex.
As the measure is modified, varying $\beta$, we observe
a soft geometric phase transition from the elongated phase to 
a crinkled phase for $\beta \lesssim -1$, in agreement with
the phase diagram Figure~1.  Moreover, a scaling analysis
of the transition is consistent with a third order
transition ($\alpha = -1$).

In the crinkled phase the fractal structure is dominated by 
a gas of sub-singular vertices, i.e.\ vertices whose order
grows sub-linearly with the volume.  The intrinsic fractal structure
is characterized by a set of critical exponents:
({\tt i}) A negative string susceptibility exponent $\gamma$, which
  decreases with $\beta$.
({\tt ii}) A Hausdorff dimension which is either $d_H = \infty$,
  or $d_H^{\prime} \approx 2$ (as $\beta \rightarrow -\infty$),
  depending on whether it is measured on the
  triangulation itself or its dual graph.
({\tt iii}) A spectral dimension, measured on the
 dual graph, which increases smoothly
 from a branched polymer value $d_s = 4/3$, at $\beta = 0$,
 to $d_s \approx 3/2$ at $\beta = -4$.

We have collected some of the more 
technical issues in two appendixes:  
In Appendix~A we list the SCE of the three different series, and in 
Appendix~B we describe some complications in extracting $\gamma$
from measurements of minbu distributions on maximal
triangulations.


\section{Maximal triangulations}

A maximal triangulation is a triangulation which for given 
volume $N_D$ has the maximal allowed number of vertices $N_0$.  
For a combinatorial triangulation this implies:
\begin{equation}
 N_0^{max} \;=\; \left \{ \begin{array}{cc}
  \left\lfloor  \frac{\textstyle N_3+10}{\textstyle 3}\right\rfloor  
     & D = 3\;,  \\ \\
   \left\lfloor \frac{\textstyle N_4+18}{\textstyle 4}\right\rfloor  
     & D = 4\;.
   \end{array} \right .
   \label{maximal}
\end{equation}
Here $\lfloor x \rfloor$ denotes the floor
function, i.e.\ the biggest integer not greater than $x$.
From this definition it follows that in general more then one volume
will correspond to a given maximal vertex number. 
In $3D$ this defines three different {\em series}, labeled by 
the $\bigl(N_0,N_3\bigr)$--pairs:
\begin{equation}
 S^0\;\bigl (N_0,3 N_0\!-\!10\bigr ), 
 \qquad S^1\;\bigl (N_0,3 N_0\!-\!9\bigr ), 
 \qquad {\rm and} \qquad S^2\;\bigl (N_0,3 N_0\!-\!8\bigr ).
\end{equation} 
The first series $S^0$ has the smallest volume 
for a given number of vertices
and we will call it the {\em minimal} series.

Likewise, in $4D$ there are in principle four different series. 
However, as only combinatorial four-triangulations of {\it even} 
volume are allowed this reduces to only two:
\begin{equation}
 S^0\;\bigl (N_0,4N_0\!-\!18\bigr ) \qquad {\rm and}  
 \qquad S^1\;\bigl (N_0,4 N_0\!-\!17\bigr ). 
\end{equation}

In the space of all triangulations, the different series are
related through a set of (topology preserving) geometric changes, 
the $\bigl(p,q\bigr)$--moves. In a $\bigl(p,q\bigr)$--move, where
$p = D+1-q$, a ($q-1$)--sub-simplex is replaced by 
its ``dual`` ($p-1$)--sub-simplex, provided no manifold constraint 
is violated.  These moves are used in the MC simulations
and are ergodic for $D \leq 4$, i.e.\ any two triangulations are 
related through a finite sequence of moves \cite{steen}.

\begin{figure}
 \epsfxsize=5.0in \centerline{ \epsfbox{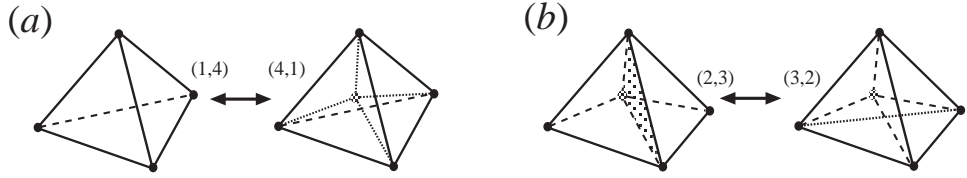}}
 \caption[movefig]{\small The ($p,q$)--moves in three dimensions:
 ({\it a}) Inserting a vertex and its inverse, deleting a vertex.
 ({\it b}) Replacing a triangle by an link and vice verse.} 
 \label{fig.move3D}
\end{figure}

In three dimensions there are two sets of moves:
Move $\bigl(1,4\bigr)$ inserts a vertex into a tetrahedra and its
inverse, move $\bigl(4,1\bigr)$, deletes a vertex of order four.   
In move $\bigl(2,3\bigr)$  a triangle separating two adjacent
tetrahedra is replaced by a link connecting the
opposite vertices.  This is shown in Figure~\ref{fig.move3D}.
These moves change the number of different $D$--simplices in 
the following way:
\begin{eqnarray}
 \begin{pmatrix}
   N_0 \\ N_1 \\ N_2 \\ N_3
 \end{pmatrix} \,
 \begin{matrix}
  {(1,4)} \\ \rightleftarrows \\ {(4,1)}
 \end{matrix} \,
 \begin{pmatrix}
 N_0+1 \\ N_1+4 \\ N_2+6 \\ N_3+3
 \end{pmatrix}
 \qquad {\rm and} \qquad
 \begin{pmatrix}
   N_0 \\ N_1 \\ N_2 \\ N_3
 \end{pmatrix} \,
 \begin{matrix}
  {(2,3)} \\ \rightleftarrows \\ {(3,2)}
 \end{matrix} \,	
 \begin{pmatrix}
  N_0 \\ N_1+1 \\ N_2+2 \\ N_3+1
 \end{pmatrix} \,.
\end{eqnarray}
Note that only the first set of moves changes both the volume
and the number of vertices, and when applied 
to a maximal triangulation we stay within the particular
series, i.e.\ the different series
are internally connected {\em via} this move.

The different series are connected, for fixed $N_0$,
by the $\bigl(2,3\bigr)$ and the $\bigl(3,2\bigr)$--moves.    
However, as it is not
possible to apply move $\bigl(3,2\bigr)$ to the series $S^0$ (as 
triangulations in this series have the minimal volume for
a given $N_0$),  the only connection which the maximal ensemble
has with the rest of the triangulation
space is {\it via} move $\bigl(2,3\bigr)$ applied to the series $S^2$. 
The relations between the different series,
{\it via} the $\bigl(p,q\bigr)$--moves, are shown in Figure~3. 
\begin{figure}
 \begin{center}
  \psfrag{m23}{$(2,3)$}
  \psfrag{m32}{$(3,2)$}
  \psfrag{m14}{$(1,4)+(4,1)$}
  \psfrag{m14r}[r][r]{$(1,4)+(4,1)$}
  \psfrag{m14c}[c][c]{$(1,4)+(4,1)$}
  \psfrag{s0}{$S^0$}
  \psfrag{s1}{$S^1$}
  \psfrag{s2}{$S^2$}
 \includegraphics[width=10cm]{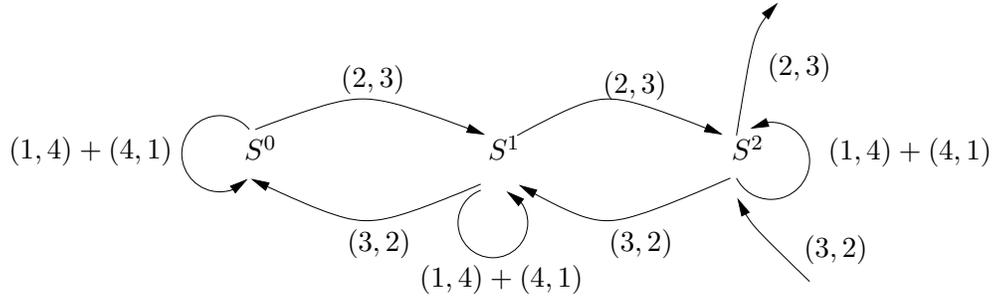}
 \end{center}
 \label{moves}
 \caption{\small The relations between the different series of
  the $3D$ maximal ensemble, $S^0$, $S^1$ and $S^2$, 
  {\it via} the $\bigl(p,q\bigr)$--moves.}
\end{figure}

\subsection{\sl Stacked spheres}

A $D$--dimensional {\em stacked sphere} \footnote{The ensemble of 
stacked spheres is as well defined in two dimensions as for $D>2$.
However, in $2D$ this does not define a maximal ensemble as 
$N_2=2N_0-4$ for every (spherical) triangulation.} 
is a triangulation constructed 
by  successivelly gluing together   smallest $D$--spheres made out 
of $D+1$, $D$--dimensional regular simplexes\cite{acm}. 
Gluing is defined as cutting away one $D$--simplex in each triangulation
and joining the two triangulations by the resulting boundary. 
Gluing a $D$--sphere corresponds to applying the 
($1,D\!+\!1$)--move.

From the definitions of stacked spheres and the maximal ensemble it 
follows that stacked spheres belong to the minimal series $S^0$.
The inverse statement, although less trivial, is also true.  
This was proven, in three and four dimensions, 
in two theorems due to Walkup \cite{w}:
\newtheorem{theorem}{Theorem}
\begin{theorem}
 For any combinatorial triangulation of a 3--sphere the inequality
 \vspace{-6pt}
 \begin{equation}
  \label{walk3}
  N_3 \,\ge\, 3N_0  - 10
 \end{equation}
 
 \vspace{-6pt}
 \noindent  
 holds with equality if and only if it is a stacked sphere.
\end{theorem}
\begin{theorem}
 For any combinatorial triangulation of a 4--sphere the inequality
 \vspace{-6pt} 
 \begin{equation}
   N_4 \,\ge\, 4N_0-18
 \end{equation}
 
 \vspace{-6pt} 
 \noindent
 holds with equality if and only if it is a stacked sphere.
\end{theorem}

From those theorems follows a non-trivial statement:
The move $\bigl(1,D+1\bigr)$, and it inverse $\bigl(D+1,1\bigr)$, 
is ergodic when applied to the ensemble of stacked spheres.  
In contrast, for the non-minimal series this statement 
is not true.  There exist triangulations, $T \in S^k$, $k > 0$,
that cannot be reduced to the minimal configuration of the
corresponding series by a repeated application of the
move $\bigl(D+\!1,1\bigr)$.  This we have verified numerically.


\section{Enumeration of  stacked spheres}

As stacked spheres have a simple tree-like structure,
it is possible to calculate the partition function 
restricted to this ensemble explicitly.   
In this section we present algorithms for doing
so, both for {\em labeled} triangulations (which
enter the MC simulations) and for the number of
{\em distinct} or {\em unlabeled} triangulations.
We also show how to generalize the algorithm for counting
labeled triangulations to
include a modified measure Eq.~(\ref{part.measure}).

All the results in this section are based on Ref.~\cite{hrs} 
which deals with the enumeration of {\em simplicial clusters}.  
A ($D+1$)--dimensional simplicial cluster is a simplicial complex 
obtained by gluing together $(D+1)$--dimensional
regular simplices along their $D$--dimensional faces. 
Comparing this to the definition of stacked spheres one sees that 
a $D$--dimensional stacked sphere is the boundary of 
a $(D+1)$--dimensional simplicial cluster where the
outer faces of the simplicial cluster correspond to the simplices of 
the stacked sphere.  So number of $(D\!+\!1)$--dimensional 
clusters with $n\;$ $(D+1)$--simplices
corresponds to the number of stacked spheres with $N_0=n+D+1$
vertices, i.e.\ $N_D=Dn+2$ simplices.

\subsection{\sl Labeled triangulations}

The basic ``building block'' is the expression for the number of
$(D+1)$--dimensional simplicial clusters build out of 
$n \; (D+1)$--simplices  rooted at a marked outer face.  
Denoting this number
by $e_{D+1,n}$ we have the following recursive relation~\cite{hrs},
\begin{equation}
 \label{enrec}
  e_{D+1,n} \;=\!\! \sum_{n_1+\cdots +n_{D+1}=n-1}
  \!\!\!e_{D+1,n_1} \cdots e_{D+1,n_{D+1}} \;.
\end{equation}
Introducing the counting series,
\begin{equation}
 \label{count}
  E_{D+1}(t) \;=\; \sum_{n=0}^\infty \; e_{D+1,n}\;t^n
            \;=\; 1 + t \bigl (E_{D+1}(t) \bigr )^{D+1},
\end{equation}
from Lagrange's inversion formula 
(see for example \cite{ri} page 147) it follows that
\begin{equation}
  e_{D+1,n} \;=\; \frac{1}{nD +1 }\;\binom{(D\!+\!1) n}{n}\;.
\end{equation}

To calculate the number of labeled simplicial clusters we note that
once the outer face has been labeled,
the remaining $N_0-D$ vertices can be labeled in $(N_0~\!-~\!D)!$
distinct ways.  The four labels on the marked face can be chosen in
one of $\binom{N_0}{D}$ ways. Including the $\frac{1}{N_0!}$ factor
from Eq.~(\ref{part.micro}), a normalization factor $(D~\!+~\!1)!$ 
(as used in Ref.~\cite{us4d}), and dividing by 
$N_D$ to adjust for the fact that we count rooted 
configurations, we arrive at the micro-canonical partition
function for stacked spheres,
\begin{equation}
 \label{wD}
  W_{S}(N_D) \;= \; \frac{D+2}{N_D} \; 
  e_{D+1,\frac{N_D-2}{D}} \;.
\end{equation}
The first 20 terms in the above series are listed in 
Tables~2 and 3 in Appendix~A for $D=3$ and 4.
This corresponds to the enumeration of all stacked spheres
of volume $N_3 \leq 62$ and $N_4 \leq 82$, respectively.

To investigate the leading behavior of the above series,
we perform an asymptotic expansion of
Eq.~(\ref{wD}) in powers of $N_D^{-1}$. 
For $D=3$ and 4 this gives:
\begin{eqnarray}
 \label{w3as}
  \qquad W_{S}(N_3) &=& \frac{10}{\sqrt{2\pi} \; N_3^{5/2}}
  \left (\frac{256}{27}\right)^{\frac{N_3-2}{3}}
  \left ( 1
     \;+\; { \frac{83}{48}\frac{1}{N_3} }
     \;+\; { \frac{11305}{4608}\frac{1}{N_3^2} }  \right. \\ 
     \nonumber \\
  && 
     \;+\; { \frac{2109275}{663552}\frac{1}{N_3^3} }
     \;+\; { \frac{503051857}{127401984}\frac{1}{N_3^4} }
     \;+\; { \frac{30688689031}{6115295232}\frac{1}{N_3^5} }
     \;+\;  \cdots  \nonumber
     \nonumber \\
     \nonumber \\\label{w4as}
  \qquad W_{S}(N_4) &=& \frac{6\sqrt{5}}{\sqrt{2\pi}\;N_4^{5/2}}
  \left (\frac{3125}{256}\right)^{\frac{N_4-2}{4}}
  \left ( 1
     \;+\; { \frac{33}{20}\frac{1}{N_4} }
     \;+\; { \frac{1729}{800}\frac{1}{N_4^2} }  \right. \\ 
     \nonumber \\
  && 
     \;+\; { \frac{201159}{80000}\frac{1}{N_4^3} }
     \;+\; { \frac{19060503}{6400000}\frac{1}{N_4^4} }
     \;+\; { \frac{556062507}{128000000}\frac{1}{N_4^5} }
     \;+\;  \cdots  \nonumber
\end{eqnarray}
In both cases, the leading asymptotic behavior is of the form
Eq.~(\ref{def.gamma}) with $\gamma=1/2$.  Moreover, 
for both series the corrections to 
the leading order are analytic in $N_D^{-1}$, i.e.\ there are no
confluent singularities.  This behavior is essential for applying
the particular variant of the ratio method we use 
to analyze the series in next section.

In contrast, the enumeration of the unlabeled triangulations 
--- the number of distinct graphs --- is much harder as the
symmetry factor of each triangulation has to be canceled.  
This is the main subject of Ref.~\cite{hrs} which provides
an algorithm to construct the generating function
\begin{eqnarray}
 \label{unlab}
  h_{D+1}(t)  \;=\; \sum_{n=0}^{\infty} \, h_{{D+1},n} \;t^n,
\end{eqnarray}
where $h_{D+1,n}$ is the number of distinct ($D$+1)--dimensional 
simplicial clusters build out of $n$ simplices.  
We list the functions $h_4(t)$ and $h_5(t)$ in Appendix~A
together with the first 20 terms (Tables~2 and 3).  
It is interesting to note that although the series of unlabeled
triangulations has the same asymptotic behavior as 
the labeled one, i.e.\ the same $\mu_c$ and $\gamma$,
the sub-leading corrections are not analytic in this
case.  This is evident from an analysis of the
singularities of the functions $h_4(t)$ and $h_5(t)$.
As a consequence, for unlabeled triangulations
it is is impossible to extract the leading behavior using 
a series extrapolation method such as the ratio method
(as is demonstrated in Figure~\ref{fig.zeta_c}).

\subsection{\sl A modified measure}

It is possible to extend the above algorithm for enumerating labeled
stacked spheres to accommodate the measure
term $q^{\;\beta}$.  The additional complication is to keep
track of the order of each of the vertices on the marked face. 
Here we present a recursive algorithm to do this.
However, for simplicity we consider only 
$3D$ stacked spheres, i.e.\ $4D$ simplicial clusters.

We denote by $e({\bf q};n,\beta)$  the sum over all triangulations
rooted at a marked outer face, where
${\bf q} = q_1,\ldots,q_4$ is the order of the
vertices at the root. Each triangulation is weighted by
a factor 
\begin{equation}
 \prod_{i=5}^{N_0} \;q_i^{\;\beta},
\end{equation}
where the product is over all vertices  {\em not} belonging to the 
marked face.  Thus the number of rooted simplicial clusters becomes
\begin{align}
 \label{enrecmes} \nonumber
 \;\;\;e({\bf q};n,\beta) \;=  \hspace{-5pt}
 \sum_{\stackrel{n_1 + n_2 + n_3 + n_4}{=n-1}}
 \; \sum_{{\bf a},{\bf b},{\bf c},{\bf d}}
  &\;\;\;e({\bf a};n_1,\beta)
   \;e({\bf b};n_2,\beta)
   \;e({\bf c};n_3,\beta)
   \;e({\bf d};n_4,\beta) \\[-14pt]
  &\hspace{30pt}
  \times\bigl (a_4+b_4+c_4+d_4-4 \bigr )^{\beta} 
  \\[4pt] \nonumber
  &\hspace{-55pt}\times\delta_{a_1 + b_1 + c_1 -2, q_1}\;
  \delta_{d_1 + a_2 + b_2 -2 , q_2} \;
  \delta_{c_2 + d_2 + a_3 -2 , q_3}\;
  \delta_{b_3 + c_3 + d_3 -2 , q_4}\;,
\end{align}
with the normalization
\begin{equation} 
 e(2,2,2,2;0,\beta) \;=\; 1.
\end{equation}
We have to sum over all possible vertex orders of the 
marked face,
\begin{eqnarray}
 e(n,\beta)&=&\sum_{\bf q} \;\bigl (q_1\, q_2\, q_3\, q_4 \bigr )^\beta 
 \;e({\bf q};n,\beta)\;,
\end{eqnarray}
to obtain the micro-canonical partition function
\begin{equation}
  W_{S}(N_3,\beta) \;=\; \frac{1}{\cal N}\;\frac{5}{N_3}
  \;e\bigl ({\scriptstyle \frac{N_3-2}{3}},\beta \bigr ),
\end{equation}
where ${\cal N}=4^{5\beta}$ is the weight of the 
smallest triangulation.

Although a solution of Eq.~(\ref{enrecmes}) is not known in 
a closed form, it is possible to calculate the series 
$W_{S}(N_3,\beta)$ recursively.  A
simple program written in Fortran 90 using quadruple
precision takes about 7 hours of CPU time on a 600MHz Alpha station to
compute $W_{S}(N_3,\beta)$ for $N_3\le 53$ $(n\le 17)$. 
Each additional term in the series requires four times the
amount of computation as the previous one.

Compared to a method of directly identifying distinct 
triangulations in MC simulations, discussed in next section, 
using the algorithm presented above it is possible to calculate 
more terms in the SCE.  However, method presented above is limited to the 
enumeration of stacked spheres, i.e.\ to a sub-class of
the maximal ensemble, and moreover the series has to
be calculated separately for each value of $\beta$. 


\section{Pure maximal triangulations}

We will start our exploration of the maximal ensemble
with the pure (unmodified) case $\beta = 0$.
As stated, the partition function Eq.~(\ref{part.max})
effectively separates into three distinct series,
$S^0$, $S^1$ and $S^2$, corresponding to regions
of the triangulation-space only loosely connected
{\it via} the ($p,q$)--moves.
In this section we would like to explore in more
details the difference between those series, in 
particular their asymptotic behavior. 

The following heuristic argument gives some
idea of what asymptotic behavior we can expect.
The series $S^1$ and $S^2$, which are close to
but do not saturate the upper bound in Walkup's theorem
Eq~(\ref{walk3}),
can be considered as stacked spheres with one, or two, 
defects, respectively. 
A defect is defined as an one application of
the move ($2,3$), i.e.\ replacing a triangle by a link.  
For a sufficiently large triangulation one naively
expects that the number of ways such a defect
can be inserted should grow linearly
with the volume 
(inserting defects into
two distinct non-symmetric stacked spheres does not lead
to identical triangulations).
This implies that although the
leading exponential behavior of the three series might be the same,
the sub-leading behavior is should be multiplied by 
$N_3$ (or $N_3^2$), hence $\gamma(S^1) = 1.5$ 
(or $\gamma(S^2) = 2.5$), with $\gamma$ defined 
in Eq.~(\ref{def.gamma}).

\begin{figure}
\epsfxsize=4in \centerline{ \epsfbox{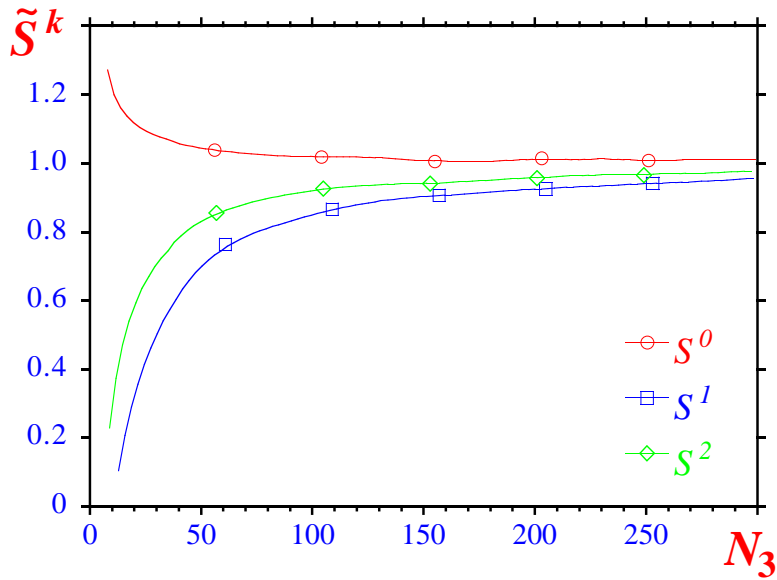}}
\caption[xxx]{\small The partition functions $\tilde{S}^k$ of
 the three distinct series of the maximal ensemble in $D=3$.
 The leading behavior, 
  ${\rm e}^{\;\mu_c(k) N_3}\;N_3^{\gamma(k)-3}$, has been
 divided out and the curves are normalized with an estimate 
 of the value at $N_3 = \infty$.
 The symbols just identify the different series.}
\label{fig.free}
\end{figure}

To investigate the asymptotic
behavior, and to verify the statement above,
we have performed several numerical experiments.
In Figure~\ref{fig.free} we show the 
micro-canonical partition function for the
three series measured numerically for $N_3 \leq 300$.
We have divided out the leading behavior of the
partition functions, i.e.\ we plot
\begin{equation}
 \widetilde{S}^k \;=\; c_k \; {\rm e}^{\textstyle -\mu_c(k) N_3}
  \;N_3^{\;3-\gamma(k)} \; S^k\;,
\end{equation}
where $c_k$ is some convenient normalization and we  
assume the exact values
\begin{equation}
 \label{eq.assump}
 \mu_c(k) \;=\; \frac{1}{3} \log \left( \frac{256}{27} \right)
 \qquad {\rm and} \qquad
 \gamma(k) \;=\; \frac{1}{2} + k\;.
\end{equation}
The behavior in Figure~\ref{fig.free} is indeed
compatible with this assumption.  
If the series had different $\mu_c(k)$ the curves 
should approach a straight line with a non-zero slope.  
Likewise, a different value of $\gamma(k)$ would manifest 
itself in a very slow convergence.  Neither behavior is observed, 
all three curves in Figure~\ref{fig.free} converge rapidly on 
a horizontal line.  It should be noted, however, that the two
non-minimal series appear to have larger finite-size
corrections.

\subsection{\sl The strong-coupling expansion}

A more precise determination of the critical parameters,
$\mu_c(k)$ and $\gamma(k)$, comes from the SCE
of the micro-canonical partition function 
Eq.~(\ref{part.max}).
In Section~3 we presented a algorithm for calculating
the SCE restricted to the minimal series $S^0$.
A corresponding recursive algorithm for 
constructing the two non-minimal series is, 
however, not available.
In fact, it is not clear that such algorithm exist.
Unlike ${\cal S}^0$, the
non-minimal series cannot be constructed from some
(different) minimal configuration by a successive
application of move ($1,4$).

We thus resort to an alternative method to enumerate
the two non-minimal series. This method, based on
a direct identification of all distinct triangulations in
MC simulations, has previously been
applied in four dimensions to identify all distinct
triangulations of volume $N_4 \leq 38$ \cite{us4d}.
The two main ingrediences of the method are:
({\sf i})
 A  sufficiently complicated hash function $f(T)$ that uniquely 
 identifies combinatorially distinct triangulations.
({\sf ii})
 A calculation of the corresponding symmetry factors, $C_T$, 
 by an explicit permutation of the vertex labels.
 
The SCE's for $S_1$ and $S_2$ are 
shown in Table~4 in Appendix~A.
As the two non-minimal series grow much faster
than the minimal one, this method is limited to
volume $N_3 \leq 36$.  This corresponds to 10 and 8
terms respectively.  To calculate the next term would
require the identification, and storing, of $\approx 3\times10^7$ 
distinct triangulations --- a formidable task with the computer
resources available.

To analyze the SCE and to extract the asymptotic behavior
we apply an appropriate series extrapolation method.
For these particular series the most powerful method is a 
variant of the ratio method, a Neville-Aitken extrapolation, 
in which successive correction terms to the partition function 
are eliminated by a suitable sequence extrapolation technique
\cite{gutmann}. 
For the minimal series this method yields impressively
accurate estimates of the critical parameters.
Applied to the first 20 terms in Table~2 we get
\begin{eqnarray}
 \nonumber
 \mu_c(0)  &= &0.749780192825077800383(15)\;,  \\
 \nonumber
 \gamma(0) &= &0.500000000000000000002(34)\;,
\end{eqnarray}
compared to the exact values $\gamma = \frac{1}{2}$ and
\begin{eqnarray}
 \nonumber
 \mu_c(0) &= &0.74978019282507780038404\ldots 
\end{eqnarray} 
The quoted errors indicate how the estimate change 
as the last term in the series is excluded.

For the non-minimal series the finite-size corrections
are larger and, in addition, we have less terms available.  
The estimates of the critical parameters are
correspondingly less accurate, especially for $S^2$:
\begin{eqnarray}
 \nonumber
 &\left. \begin{array}{lll}
           \mu_c(1)  &= &0.749765(17)  \\
           \gamma(1) &= &1.497(18)
	\end{array}  \right \} &S^1\;, \\ \nonumber
 &\left. \begin{array}{lll}
           \mu_c(2)  &= &0.74940(49)  \\
           \gamma(2) &= &2.24(30)
	\end{array}  \right \} &S^2\;.
\end{eqnarray}
These values are though consistent with the conjecture
that $\mu_c$ is the same for all the three series, but that 
$\gamma(k) = \frac{1}{2} + k$.

The same behavior is observed for the two series of the
maximal ensemble in $4D$. The first 20 terms in Table~3 yield 
equally impressive agreement with the exact values:
$\mu_c = \log \sqrt[4]{3125/256} \approx 0.625503\ldots$
and  $\gamma = \frac{1}{2}$.
And the first 7 terms in the non-minimal series $S^1$,
given in Ref.~\cite{us4d}, give $\mu_c(1) = 0.6245(17)$
and $\gamma(1) = 1.64(35)$. 

Some final comments regarding the
non-minimal series.  Although an exact enumeration
algorithm is not available it is possible to obtain a lower 
bound on the series $S^1$ and $S^2$ by assuming that all
non-minimal triangulations can be constructed by move 
$\bigl(1,4\bigr)$ from the appropriate minimal configuration.
The resulting SCE has, for both series\footnote{Note that for
the series $S^2$ this assumes that the two defects are 
connected, thus excludes triangulations with 
two randomly distributed defects.  This results in a lower
estimate $\gamma = \frac{3}{2}$, instead of $\gamma = \frac{5}{2}$
which we expect for $S^2$.}, the same $\mu_c$ and
$\gamma = \frac{3}{2}$.

\subsection{\sl Distributions of baby universes}

The different asymptotic behavior 
for the three series is very pronounced
when one considers
the distribution of minimal neck baby universes (minbu's)
measured on maximal triangulations.  
A minbu is a part of a triangulation connected to
the rest {\it via} a minimal neck --- on a three dimensional
combinatorial triangulation a minimal neck consists
of four triangles glued along their edges to form a
closed surface (a tetrahedra).  By counting in how many ways
a ``baby'' of size $b$ can be connected to a ``mother''  
of size $N_3 - b +2$, the distribution of minbu's can
be expressed in terms of the canonical partition functions
of the corresponding volumes \cite{minbu1}.

For a maximal triangulation the situation is more
complicated as the baby, the mother, and the ``universe'' 
can belong to any one of the different series $S^k$.  
It is not even obvious that
if a maximal triangulation is split along a minimal neck
the two parts belong to the maximal ensemble at all.
This is though easy to verify\footnote{The converse is 
however not true.  If we glue together
two maximal triangulations, $T_1 \in S^{k_1}$ and $T_2 \in S^{k_2}$, 
the result only belongs to the maximal ensemble provided 
$k_1+k_2 \leq 2$.}.
In general the distribution of baby universes 
measured on maximal triangulations becomes
\begin{eqnarray}
 \label{babymax}
 \nonumber
{\cal B}^{k_B,k_M,k_U}(b) &=&{\cal C}_{N_3}(b) \,
 \frac{b \; S^{k_B}(b) \;(N_3-b+2)
 \;S^{k_M}(N_3-b+2)}{S^{k_U}(N_3)} \\[2pt]
  &\sim & b^{\gamma(k_B)-2} \; (N_3-b+2)^{\gamma(k_M)-2}\;.
\end{eqnarray}
The indices $k_B$, $k_M$, and $k_U$ refer to the baby,
the mother, and the universe, the factors $b$ and $N_3-b+2$ count the 
possible locations of the neck, and  ${\cal C}_{N_3}(b)$ 
is a {\it contact} term due to interactions on the neck.  
The contact term is discussed further in Appendix~B; 
for pure simplicial gravity this is just a trivial constant.
In the last step we have used the expected asymptotic
behavior of the series Eq.~(\ref{def.gamma}).

\begin{figure}
\epsfxsize=4in \centerline{ \epsfbox{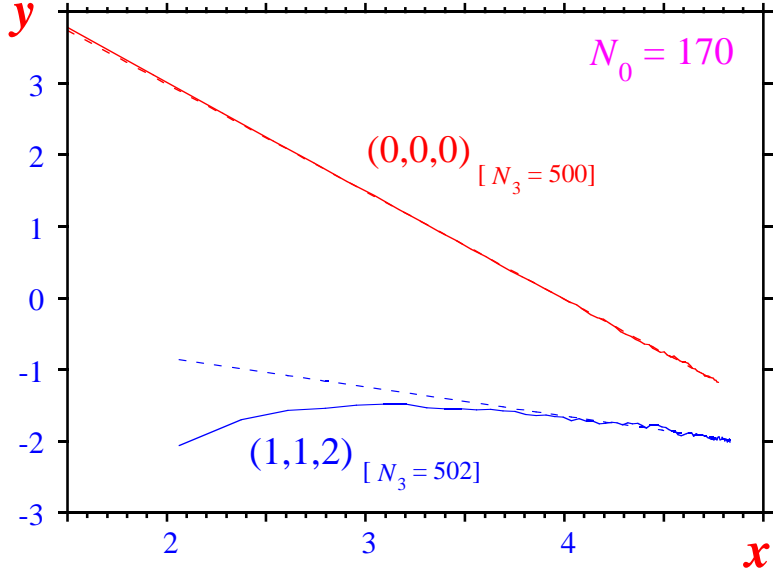}}
\caption[tub_scal]{\small The minbu distributions,
 ${\cal B}^{0,0,0}$ and ${\cal B}^{1,1,2}$,
 measured on triangulations of size $N_0 = 170$
 (i.e.\ $N_3 = 500$ or 502). In the figure we plot
 $y = \log {\cal B}$ {\rm vs}.\ 
 $x = \log \bigl (b\;(1-b/N_3) \bigr )$, and the
 dashed lines are fits to Eq.~(\ref{babymax}).}
\label{fig.baby3D}
\end{figure}

In principle, Eq.~(\ref{babymax}) defines nine different 
distributions, however the additional restriction, 
$k_B + k_M \leq 2$, reduces the number of possible combinations
to six. Two of those, ${\cal B}^{0,0,0}$ and ${\cal B}^{1,1,2}$, 
both corresponding to $k_B = k_M$, are plotted in Figure~\ref{fig.baby3D}.
Both distributions are measured on triangulations
with $N_0 = 170$, but on different volumes,
$N_3=500$, and 502.  
The former distribution, ${\cal B}^{0,0,0}$, which is composed
solely of the minimal series, has remarkably small finite-size 
effects.  A fit to Eq.~(\ref{babymax}),
including minbu's $b \geq 10$, gives $\gamma(0) = 0.495(4)$.
In contrast, for an acceptable fit to the distribution 
${\cal B}^{1,1,2}$ a lower cut-off, $b > 50$, had to be imposed.
The result, $\gamma(1) = 1.44(4)$, is consistent with our 
previous estimate.

The remaining four distributions in Eq.~(\ref{babymax})
correspond to a baby and a mother belonging to 
different series.
However, the mixture of different $\gamma$'s and 
large finite-size effects makes a reliable estimate of 
$\gamma(k)$ very difficult from these distributions.


\section{A modified measure}

We now turn to the maximal ensemble
modified by varying the measure Eq.~(\ref{part.measure})
for $\;-4 \leq \beta \leq 2$.   
For most part we restricted our investigation to
the ensemble of stacked spheres where we expect
less finite-size effects.  This provides the additional
benefit that the MC simulations are much simplified
as only move $\bigl(1,4\bigr)$, and its inverse, are required
for the updating process to be ergodic.   
However, we also did extensive simulations
of the full model (including finite values of $\kappa_0$)
to verify that this did not introduce any bias in the
measurement process.  Typically we collected about
10 to 50 thousand independent measurements at
each value of $\beta$ and on different volumes
$N_3 = 125$ to 8000.

\subsection{\sl Evidence for a phase transition}

As the fluctuations in the vertex orders are enhanced
by modifying the measure, i.e.\ by decreasing $\beta$,
the internal geometry changes.  The polymerization
of the triangulations is suppressed and the model enters
a new phase akin to the crinkled phase observed in
four dimensions \cite{us4d}.  The nature of the internal
geometry in the crinkled phase is discussed in  Section~5.2, 
here we present what evidence we have for the existence of a 
continuous phase transition at a critical value $\beta_c$.

\begin{figure}
\epsfxsize=4in \centerline{ \epsfbox{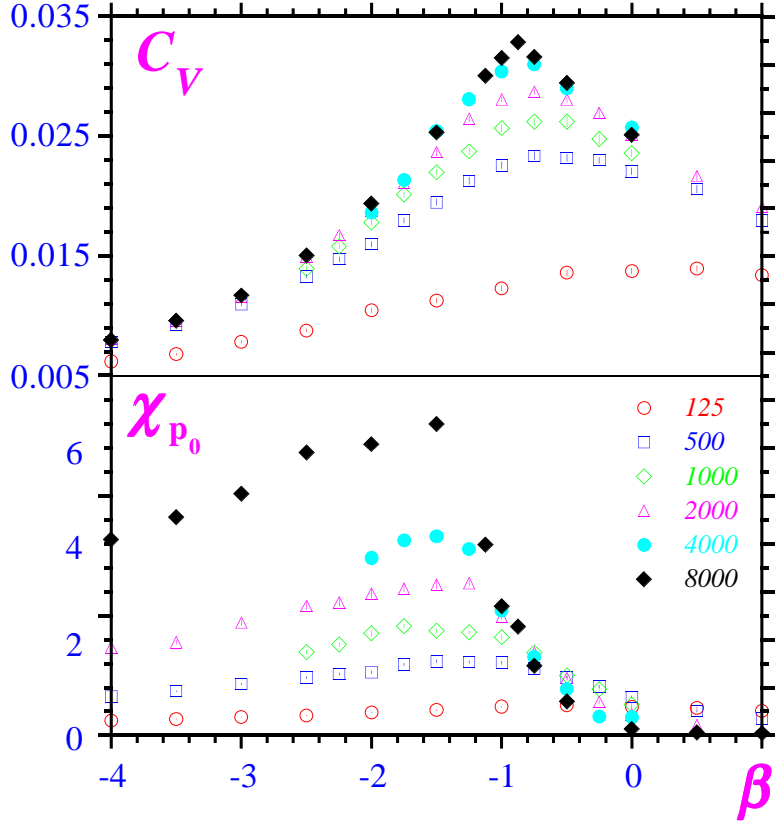}}
\caption{\small Evidence for a phase transitions as
 the measure is modified varying $\beta$.
 ({\sf Top}) The fluctuations in the energy density,
 $C_V = N_3 \bigl(\langle e^2\rangle - \langle e \rangle^2\bigr)$,
 where $e = N_3^{-1}\sum_i \log (q_i)$.
 ({\sf Bottom})~The fluctuations in the maximal vertex
 order, $\chi_{q_0} =  N_3 \bigl(\langle e q_0\rangle 
  - \langle e \rangle \langle q_0 \rangle\bigr)$.}
\label{fig.trans}
\end{figure}

\begin{figure}[t]
 \epsfxsize=4in \centerline{ \epsfbox{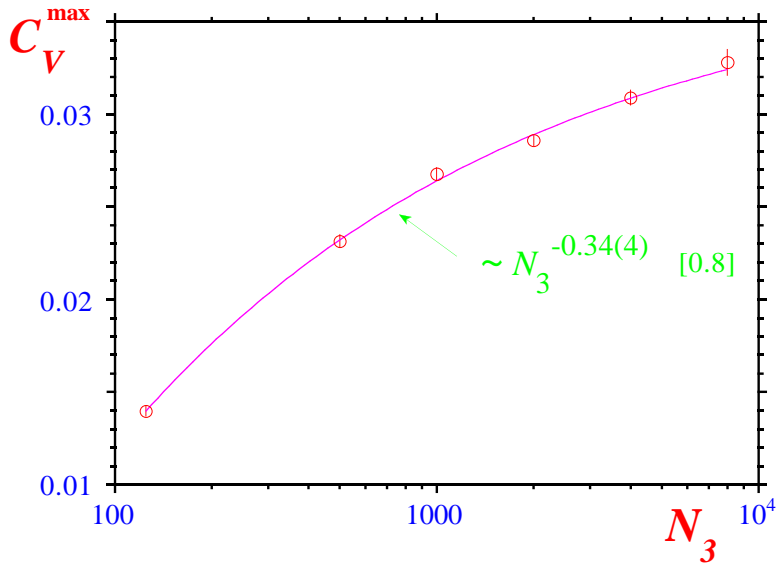}}
  \caption{\small The volume scaling of the peak in the
  energy fluctuations, $C_V^{max}$.  The curve is a 
  fit to the scaling form: $a + bN_3^x$; the optimal fit yields
  $x = \alpha/\nu d_H = -0.34(4)$ with a 
  $\chi^2/({\rm d.o.f.}) \approx 0.8$}
 \label{fig.cvmax}
\end{figure}

In Figure~\ref{fig.trans}{\it a} we show the fluctuations in the 
energy density, 
$C_V =N_3 \bigl(\langle e^2\rangle - \langle e \rangle^2\bigr)$,
as we vary $\beta$.  This is for $N_3 =125$ to 8000
and we define $e = N_3^{-1}\sum_i \log (q_i)$.
We observe on all volumes a peak in $C_V$
consistent with the existence of a phase transitions
at $\beta \approx -1$.  The peak value appears to 
saturate as the volume is increased; this makes it difficult
to distinguish between a soft continuous phase transition and 
a cross-over behavior.  The scaling of the peak value is, however,
consistent with a third order phase transition.  This is shown
in Figure~\ref{fig.cvmax} where we fit the maximal value to the
scaling behavior: $C_V^{max} \approx a + bN_3^{\alpha/\nu d_H}$.
The fit yields $\alpha/\nu d_H = -0.34(4)$, with a 
$\chi^2/({\rm d.o.f.}) \approx 0.8$.  Assuming hyper-scaling,
$\alpha = 2 - \nu d_H$, is valid this implies a
specific heat exponent $\alpha = -1.03(9)$.

Additional evidence for a transition is provided by other
geometric observables such as the maximal vertex order
$q_0$ whose value jumps at the transition.  As discussed in
the Introduction, and is quantified in the Section~5.2,
the crinkled phase is characterized by a gas of sub-singular
vertices, i.e.\ vertices whose local volume grows like
$N_3^x$, $x < 1$.  As a consequence both the fluctuations in
the maximal vertex order, 
$\chi_{q_0} = N_3 \bigl (\langle q_0^2 \rangle - 
\langle q_0 \rangle^2 \bigr )$,
and its energy derivative, 
${\rm D}_{q_0} = N_3 \bigl (\langle q_0 e \rangle 
- \langle q_0 \rangle \langle e \rangle \bigr )$, 
diverge at the transition.  
An example of the former is shown in Figure~\ref{fig.trans}{\it b}.
In fact, the susceptibility $\chi_{q_0}$ appears to diverge
in the whole crinkled phase in contrast to its behavior
in the elongated phase.

Note that some of the data in Figure~\ref{fig.trans} are
measured on non-minimal triangulations (e.g.\ $N_3 = 1000$ 
and 4000).  This shows that the phase transition is a generic
behavior of the maximal ensemble, not just of the minimal
series.  The same is true for the nature of the crinkled
phase explored in the next section.

\subsection{\sl The crinkled phase}

To investigate the nature of the internal geometry
in the crinkled phase, and to compare it to the elongated phase, 
we have measured several aspects of the fractal structure:
the distribution of vertex orders $\pi(q)$, 
the string susceptibility exponent $\gamma$,
and both the fractal and spectral dimensions $d_H$ and $d_s$.

\subsubsection {\it The vertex order distribution}

The shape of the  vertex order distribution $\pi(q)$ is a
good indicator for the nature of the internal geometry
in the different phases.  In the elongated phase the
vertex orders are more or less equdistributed and the
maximal vertex order grows logarithmically with the
system size, $q_0 \sim \log N_3$. In the crinkled 
phase, on the other hand, vertices of large order are favored 
and the maximal vertex order grows sub-linearly, 
$q_0 \sim N_3^{\alpha}$.    
Estimates of $\alpha$, from finite-size scaling including
volume $N_3 = 125$ to 8000, indicate that $\alpha \approx 0.8$
for sufficiently large negative $\beta$.

\begin{figure}
 \epsfxsize=4in \centerline{ \epsfbox{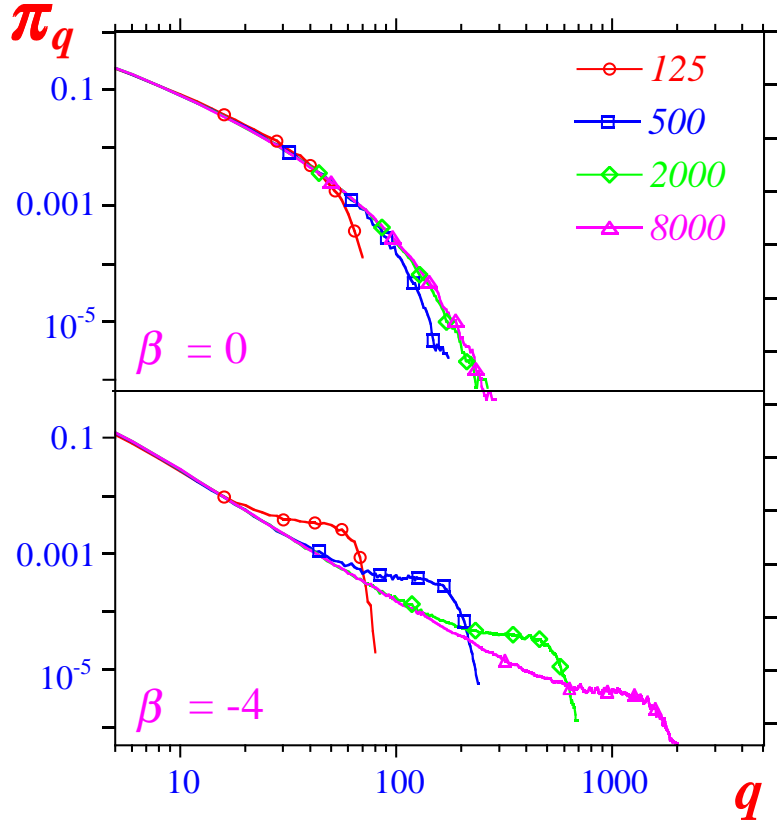}}
 \caption{\small A log-log plot of the (normalized) vertex 
  order distribution, $\pi(q)$, both in the elongated
  phase, $\beta = 0$ ({\sf Top}),
  and in the crinkled phase, $\beta = -4$ ({\sf Bottom}),
  and measured on different volumes.}
 \label{fig.vord}
\end{figure}

Examples of the  vertex order distribution in the two phases
are shown in Figure~\ref{fig.vord} for
$\beta =0$ and $\beta = -4$.
In the crinkled phase the distribution $\pi(q)$ 
developes a ``bump'' in the tail, corresponding to a
condensation of vertices of large order.  
The bump persists as the volume is increased, i.e.\ 
it captures a finite fraction of the volume as 
$N_3 \rightarrow \infty$.
This suggests that the crinkled phase can be
characterized as a {\it gas} of sub-singular 
vertices.\footnote{Similar
behavior, although more pronounced, 
is observed in the crumpled phase in $4D$ when the ensemble of 
triangulations in Eq.~(\ref{part.micro}) includes 
degenerate triangulations \cite{deg4d}.
However, in $4D$ the bump is disconnected from the rest of 
the distribution --- this justifies the nomenclature 
{\it singular structure}.}

\subsubsection{\it The string susceptibility exponent $\gamma$}

To measure the string susceptibility exponent for 
various values of $\beta$, and restricted to
the minimal series, we have used several methods:

\begin{mydescription}

\vspace{-3pt}
\item[({\sf a})]
{\sl The pseudo-critical cosmological constant}, $\mu_c(N_3)$ \\  
Assuming the asymptotic behavior Eq.~(\ref{def.gamma}), 
a saddle-point approximation of the grand-canonical 
partition function gives:
\vspace{-8pt}
\begin{equation}
 \label{eq.mu_pseudo}
 \mu_c(N_3) \;\approx\; \mu_c + \frac{\gamma-3}{N_3}\;.
\end{equation}

\vspace{-8pt}
\noindent
By measuring $\mu_c(N_3)$ with high accuracy on volumes
$N_3 \approx 100$ to 2400, and for $\beta = -1$, -2, -3,
and -4, we determined $\gamma$ by
a fit to Eq.~(\ref{eq.mu_pseudo}).

\vspace{-3pt}
\item[({\sf b})]
{\sl The minbu distribution}, ${\cal B}^{0,0,0}(b)$ \\
We measured the minbu distribution
on volume $N_3 = 2000$ and for the same values of $\beta$.
However, in contrast to the elongated phase, in the crinkled
phase the contact term ${\cal C}_{N_3}(b)$ has a non-trivial 
dependence on $b$ and has to be eliminated prior to 
extracting $\gamma$ from the (corrected) distribution
$\widetilde{\cal B}^{0,0,0}$ by a fit to Eq.~(\ref{babymax}).

\vspace{-3pt}
\item[({\sf c})]
{\sl The strong-coupling expansion} I \\
Using the enumeration methods for a non-zero $\beta$,
presented in Section~3.2, we calculated the first 17 terms of 
the SCE of stacked spheres for few values of $\beta$.  
From this SCE we extracted $\gamma$ using the ratio method.  

\vspace{-3pt}
\item[({\sf d})]
{\sl The strong-coupling expansion} II \\
By explicitly identifying all distinct  
triangulations of volume $N_3 \leq 38$, 
and calculating the corresponding measure term,
we calculate the first 12 terms in the SCE
for an arbitrary value of $\beta$.

\end{mydescription}

\setlength{\tabcolsep}{8pt}
\begin{table}
 \label{tab.gamma}
 \caption{\small Estimates of the exponent $\gamma$, for
  $\beta = -1$, $-2$, $-3$, and $-4$;
   ({\sf a}) from a fit to $\mu_c(N_3)$, Eq.~(\ref{eq.mu_pseudo}),
             (with [$\chi^2/({\rm d.o.f.})$]);
   ({\sf b}) from the minbu distribution, Eq.~(\ref{babymax}),
   both including and excluding the contact term, 
   ${\cal B}^{0,0,0}$ and $\widetilde{\cal B}^{0,0,0}$; and
   ({\sf c}) from the first 17 terms in the SCE. 
   }
 {\small
 \begin{center}
 \begin{tabular}{|cl|llll|}  \hline
 \vspace{-6pt} & & & & & \\
 &&  {\large $\beta = -1$}  &  {\large $\beta = -2$}  
  &  {\large $\beta = -3$}  &  {\large $\beta = -4$} \\  
 \vspace{-6pt} & & & & & \\ \hline
 \vspace{-6pt} & & & & & \\ 
 ({\sf a}) &$\mu_c(N_3)$
   &  -0.37(4) [1.5]  & -2.56(3) [0.8]  
   &  -5.25(2) [0.6]  & -8.11(4) [0.7]   \\ 
 \vspace{-6pt} & & & & & \\ \hline
 \vspace{-6pt} & & & & & \\ 
  ({\sf b}) & ${\cal B}^{0,0,0}(b)$
   &   $\;0.25(7)$ & -0.20(3) & -0.37(4) & -0.52(8)  \\ 
            & $\widetilde{\cal B}^{0,0,0}(b)$
   &  -0.14(5) & -2.58(7) &  -6.5(2) & -10.5(8)   \\ 
 \vspace{-6pt} & & & & & \\ \hline
 \vspace{-6pt} & & & & & \\ 
  ({\sf c}) & SCE
   &  -0.076(23) & -2.401(17) &  -5.484(42) & -8.455(23)  \\[4pt]\hline 
 \end{tabular}
 \end{center}
 }
\end{table}

\begin{figure}
 \epsfxsize=4in \centerline{ \epsfbox{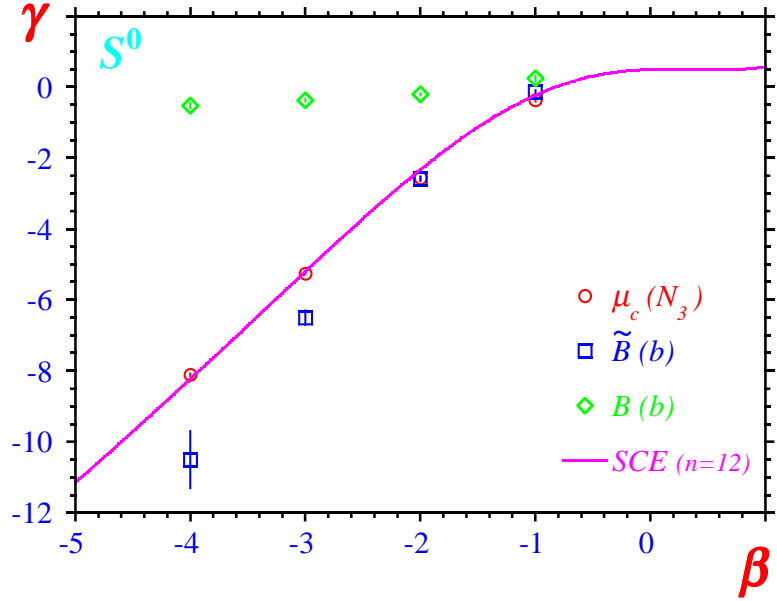}}
 \caption{\small The variation in $\gamma$ with $\beta$, 
  determined by the different methods in Table~1, This is for the
  minimal series and a modified measure Eq.~(\ref{part.measure}).
  The solid line is from the ratio method applied to
  the first 12 terms in the SCE.}
 \label{fig.gam_b00}
\end{figure}

\noindent
The estimates of $\gamma$ are listed
in Table~1 and plotted in Figure~\ref{fig.gam_b00}.  
Qualitatively, 
the agreement between the different methods is very good.
Only the estimate from the (corrected) minbu distribution
$\widetilde{\cal B}^{0,0,0}$ deviates for
large negative $\beta$.  This is understandable as 
for large negative $\beta$ the tail of the 
minbu distribution is suppressed and is difficult to measure 
accurately.

Note that if we do not correct the minbu distribution 
for the contact term ${\cal C}_{N_3}(b)$, for $\beta < 0$
a fit to ${\cal B}^{0,0,0}$ yields a very different,
and unreliable, estimate of $\gamma$.  This estimate
in included in Table~1 and Figure~\ref{fig.gam_b00}.
Why the contact term is so important for maximal
triangulations in the crinkled phase, and how it can
be eliminated, is discussed in Appendix~B.

\begin{figure}
 \epsfxsize=4in \centerline{ \epsfbox{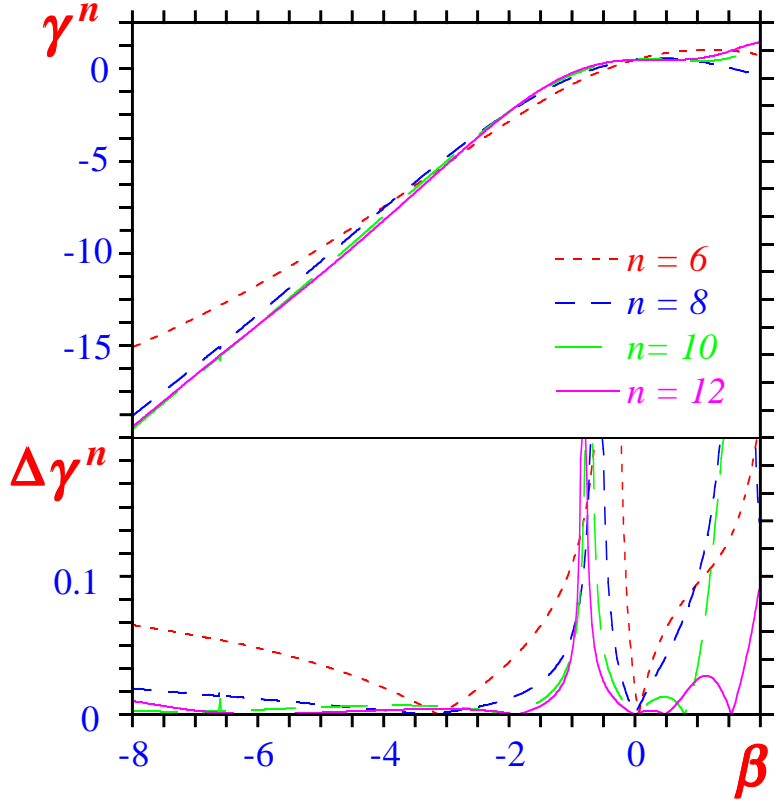}}
 \caption[tub_scal]{\small ({\sf Top}) The exponent
  $\gamma$ {\it vs}.\ $\beta$, determined from the
  first $n$ terms in the SCE using the ratio method.
  ({\sf Bottom}) The corresponding convergence of the
  ratio method, Eq.(\ref{conv}).}
 \label{fig.gamDgam}
\end{figure} 

The agreement between the different methods 
strongly supports the assumption that the asymptotic 
behavior of the canonical partition function, Eq.~(\ref{def.gamma}),
is as valid in the crinkled phase as in the elongated phase.  
There are, however, indications from the SCE that this might 
not be true at the phase transition.  This we  show in  
Figure~\ref{fig.gamDgam}.  In the upper part of the
figure we show the variation of $\gamma$ with $\beta$, estimated 
from the first $n$ terms in the SCE, for $n=6$, 8, 10 and 12.
In the lower part we plot an estimate of the convergence
of the ratio method,
\begin{equation}
 \label{conv}
 \Delta \gamma^n \;=\; \frac{1}{2}\;
  \frac{ | \gamma^n - \gamma^{n-1} | }
  {|\gamma^n|+|\gamma^{n-1}|} \,.
\end{equation}
In the elongated phase and in the crinkled phase
the estimates of $\gamma$ converge.  
However, both for $\beta \approx -1$,
and for $\beta \gg 0$, the method fails to converge.
It is possible that at the phase transition there
are some different non-analytic finite-volume 
corrections to the asymptotic behavior Eq.~(\ref{def.gamma}).
There are known examples of such behavior:
For one scalar field ($c=1$) coupled to $2D$--gravity there
are logarithmic corrections to the free energy \cite{2Dc1}.

Why the method appears to fail for $\beta \gg 0$ is less clear.
As there is no phase transition in this region,
we would expect Eq.~(\ref{def.gamma}) to be equally valid for
$\beta > 0$ as for $\beta = 0$. We note, however, that
in general the magnitude of the finite-size corrections
increases drastically for large positive $\beta$.

All the above estimates of $\gamma$ are 
for the minimal series.  We expect, however, that
the arguments presented in Section~4 
are also valid for non-zero values of $\beta$, and
that for a given $\beta$, $\gamma$ is correspondingly
larger for $S^1$ and $S^2$ than for $S^0$.
The determination of $\gamma$ is though more 
difficult in this case: 
we have no algorithm for enumerating the
non-minimal series; the minbu distributions are
more complicated (as is discussed in Section~4.2); and
as the MC simulations are not ergodic 
restricted to either of the non-minimal series, 
we cannot determine $\mu_c(N_3)$ numerically.

We are thus left only with method ({\sf d}).
As for $S^0$, we have identified all distinct
non-minimal triangulations of volume $N_3 \leq 36$;
re-weighted with the measure this yields the
variations of $\gamma(k)$ with $\beta$.
For example,  for $\beta = -2$ we get
$\gamma(S^1) = -1.2508(36)$ and $\gamma(S^2) = -0.15(33)$,
compared to $\gamma(S^0) = -2.401(17)$.  Similar results
are obtained for other values of $\beta$.
In all cases the estimates of $\gamma(k)$ are consistent with our 
expectations of how the three series behave.

\subsubsection{\sl The Hausdorff and the spectral dimension}

We have estimated both the Hausdorff and the 
spectral dimension of the maximal ensemble
in the elongated and in the crinkled phase.  
On smooth regular manifolds those two definitions of
dimensionality coincided, however
on highly fractal random geometries, likes the ones that dominate
the partition function Eq.~(\ref{part.max}), they
are in general different.

The Hausdorff dimension, $d_H$, is defined
from volume of space within a ball of geodesic radius
$r$ from a marked point: $V(r) \sim r^{d_H}$.
We have measured this both on the dual graph, from
the simplex-simplex distribution $s_{N_3}(r)$, and on
the triangulation itself, from the vertex-vertex distribution 
$v_{N_0}(r)$.
These distributions count the number of simplices 
(vertices) within a geodesic distance $r$ from a marked 
simplex (vertex).  Assuming that the only relevant 
length-scale in the model is defined by $x = N_3^{-1/d_H}$, 
general scaling arguments \cite{dh_scal} imply 
\begin{equation} 
 \label{scald_H}
 s_{N_3}(r) \;=\; N_3^{1-1/d_H}\;F(x)\;,
\end{equation}
and likewise for $v_{N_0}(r)$.  By comparing distributions
measured on different volumes, Eq.~(\ref{scald_H}) provides
an estimate of $d_H$.

\begin{figure}
 \epsfxsize=4in \centerline{ \epsfbox{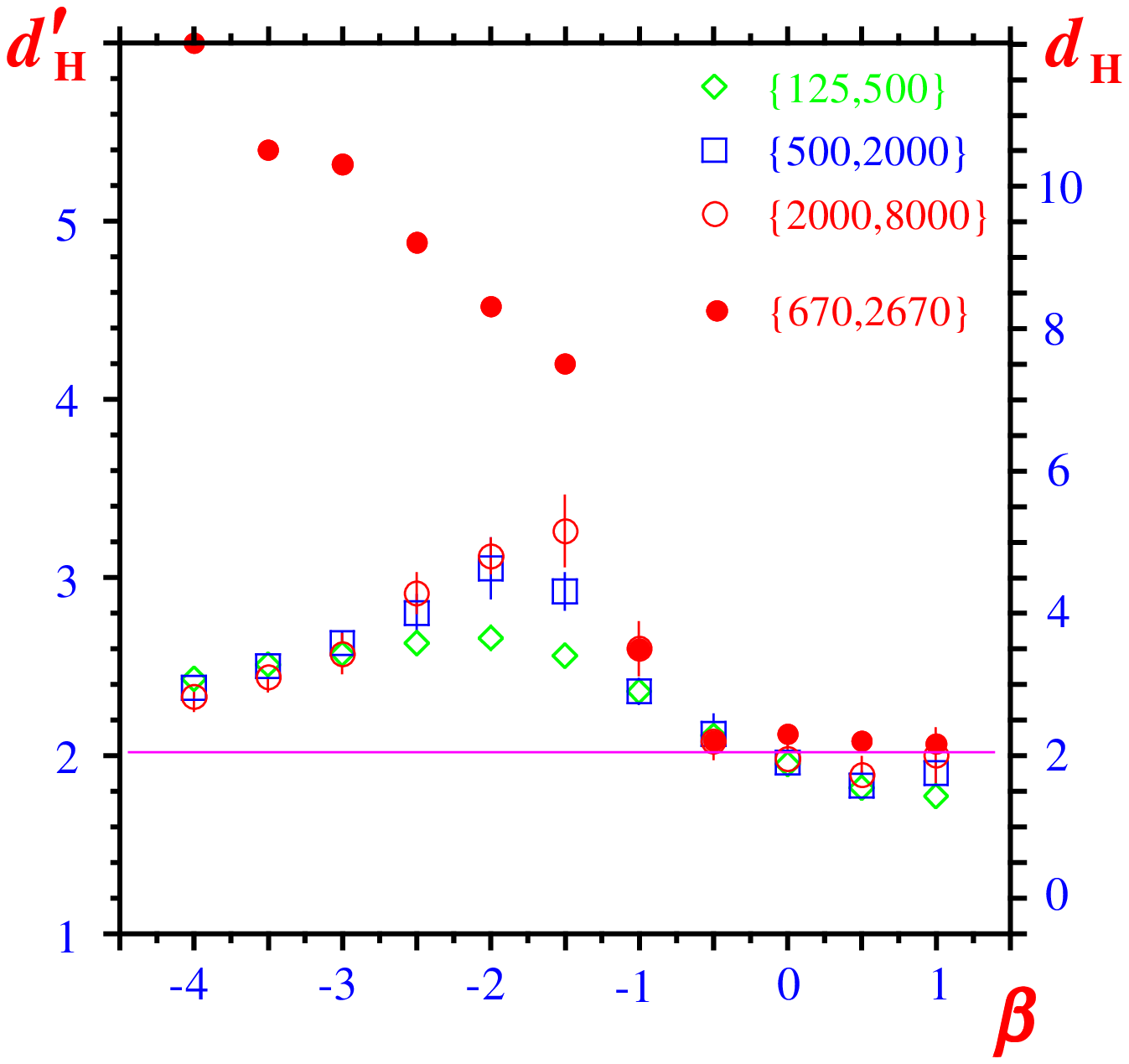}}
 \epsfxsize=4in \centerline{ \epsfbox{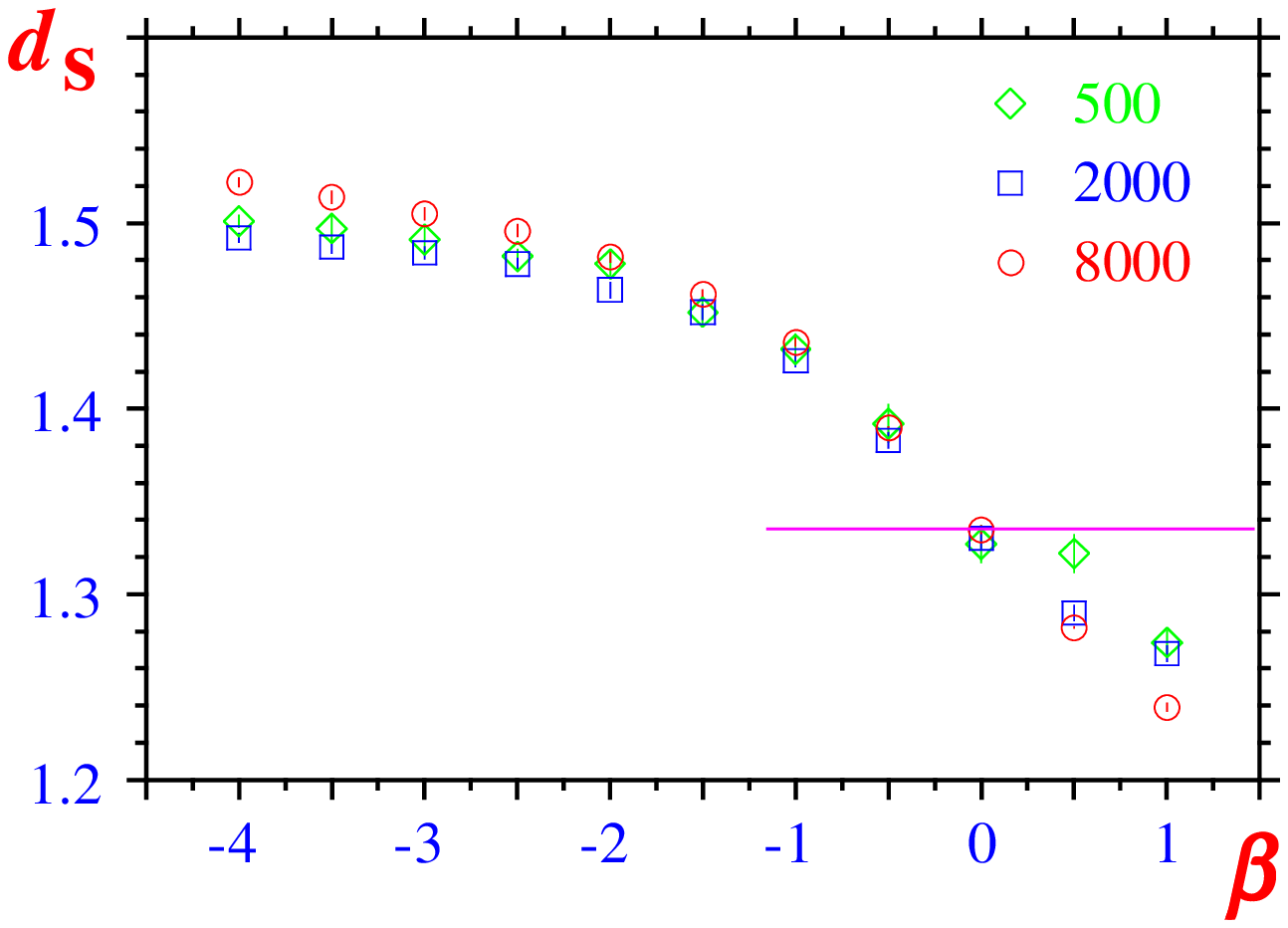}}
 \caption{\small ({\sf Top}) The Hausdorff dimension
  defined either by the 
  ({\sf a}) the simplex-simplex distribution $s(r)$, or
  ({\sf b}) the vertex-vertex distribution $v(r)$,
  $d_H^{\prime}$ (open symbols) or $d_H$ (filled symbols).
  Note that the scale for $d_H$ is double compared to
  $d_H^{\prime}$.  Triangulations of size $\{N_3,4N_3\}$
  (or $\{N_0,4N_0\}$) are used in the scaling Eq.~(\ref{scald_H}).
  ({\sf Bottom}) The spectral dimension $d_s$ measured on the
  dual graph.  In both plots the horizontal lines is the
  generic branched polymer value.}
 \label{fig.dh_pap}
\end{figure}

In principle, the Hausdorff dimension defined on
the dual lattice, $d_H^{\prime}$, need not to agree
with the one defined by $v_{N_0}(r)$, $d_H$. 
However, they do so in most models of dynamical 
triangulations which have a continuum interpretation 
as theories of extended manifolds.  
This is also the case in the elongated phase where
we expect $d_H = 2$, characteristic of generic branched 
polymers.  We measured the above distributions on
triangulations of volume $N_3 =125$, 500, 2000 and 8000,
and determined $d_H$ from the scaling Eq.~(\ref{scald_H}).
The estimates are plotted in Figure~\ref{fig.dh_pap}. 

But, not unexpected, as $\beta$ is decreased and we pass through 
the phase transition the Hausdorff dimension changes.  
What is surprising though is that in the crinkled
phase the estimates of $d_H^{\prime}$ and $d_H$ differ 
substantially.  The former, measured on the dual graph,
jumps to $d_H^{\prime}\approx 3$ at the transition,
then decreases with $\beta$, possible to $d_H^{\prime} \approx 2$ 
as $\beta \rightarrow - \infty$.  In contrast, the vertex-vertex 
distribution yields a very large estimate,
$d_H \approx 10$ for $\beta < -2$.   Such
large estimates on a finite volume suggest that the actual 
value is infinite.

This discrepancy is much too large to be explained by an
uncertainty in the determination.  It is correct that
in the crinkled phase the vertex-vertex distribution 
has an extends of only few geodesic steps\footnote{As a consequence,
only the two largest volumes, corresponding to $N_0 = 670$ and
2670, are used in determining $d_H$.  And the volume scaling
was not sufficiently good for a 
reliable error estimate.} 
(all vertexes are very ``close''), 
even on volume 8000. This makes the determination of $d_H$ 
correspondingly difficult.  
However, this is just a signal of an infinite fractal dimensions..   
The determination of $d_H^{\prime}$ from $s(r)$ is, on the
other hand, very reliable and, if anything, the estimates
appear to decrease with volume.  We show examples of both
distributions in Figure~12 for $\beta = -4$.
Both distributions have been scaled 
according to Eq.~(\ref{scald_H}) using the corresponding
optimal values of $d_H$.

\begin{figure}
 \label{labb}
 \begin{center}
  \mbox{\psfig{figure=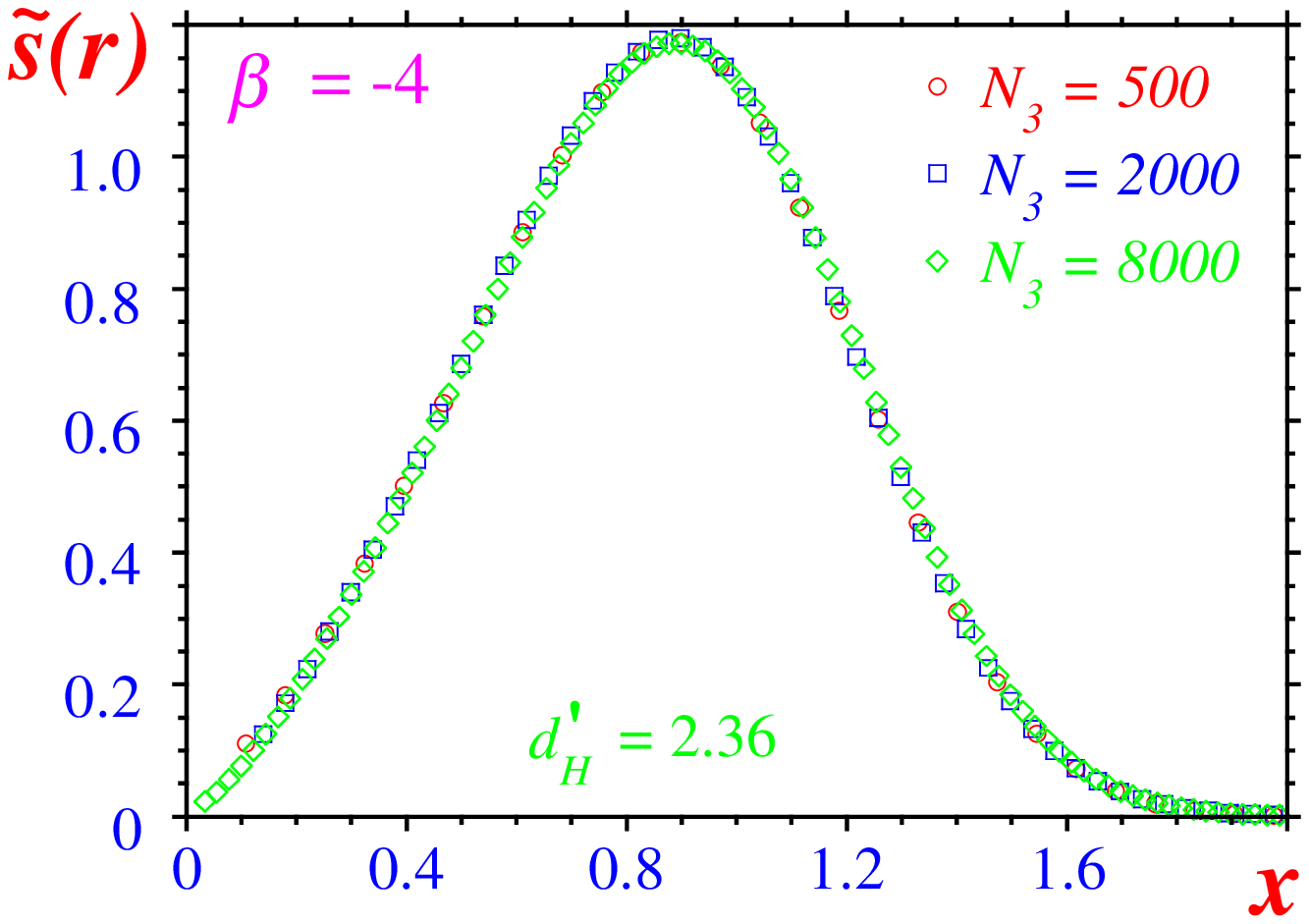,width=6.50cm} \hspace{0.2in} 
        \psfig{figure=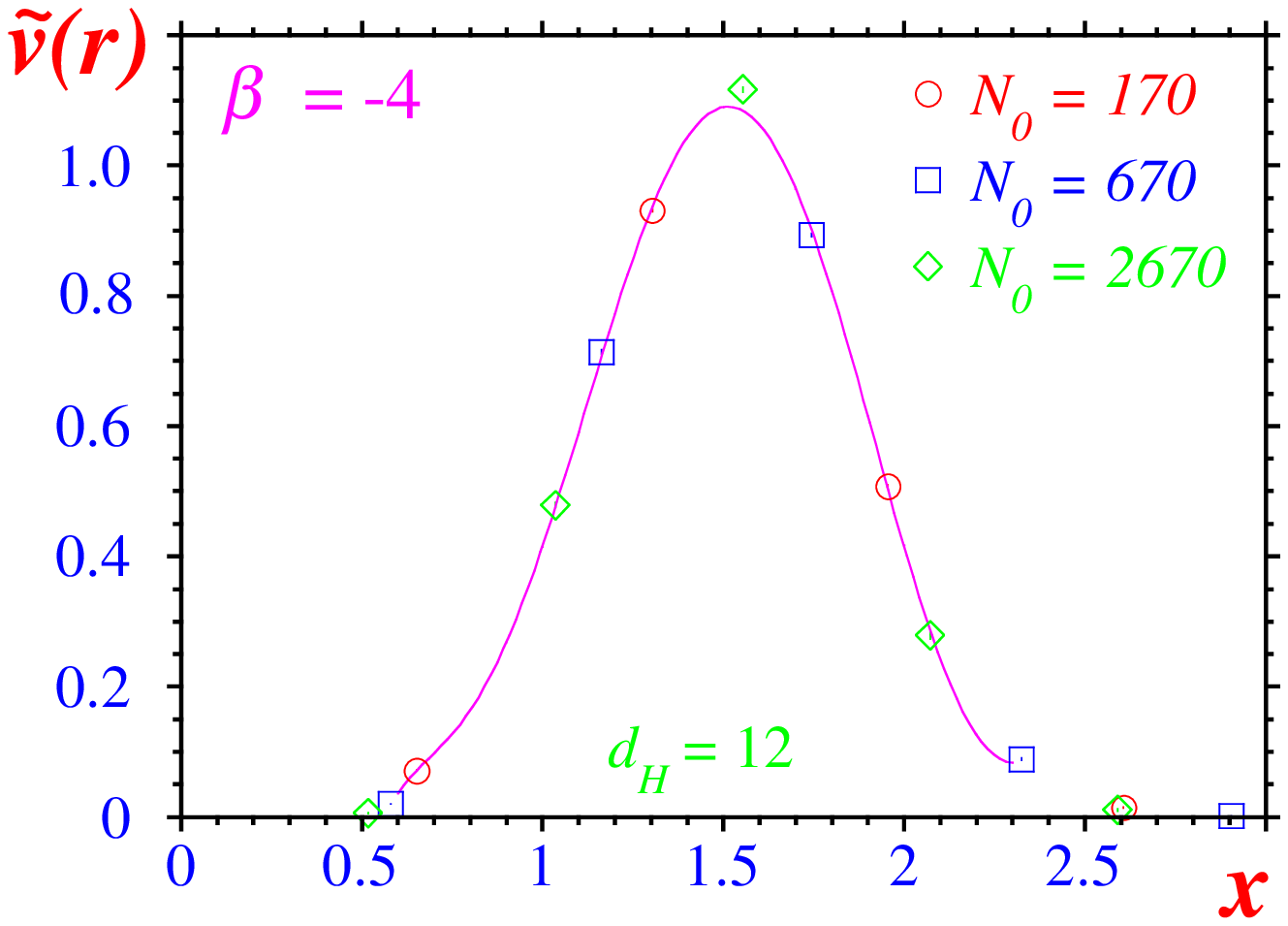,width=6.50cm} }     
 \end{center}
 \caption[labb]{\small ({\sf Left}) The simplex-simplex and ({\sf Right})
  the  vertex-vertex distributions, $s(r)$ and $v(r)$, for $\beta = -4$
  and $N_3 = 500$, 2000 and 8000.  Both distributions are
  scaled according to Eq.~(\ref{scald_H}) and are plotted {\it vs}.\
  the scaling variable $x = r/N_3^{1/d_H}$.
  Notice, however, that different fractal dimensions,
  $d_H^{\prime} = 2.36$ and $d_H = 12$, are used
  in scaling the two distributions.  The solid line is a polynomial
  interpolations.}
\end{figure}

Finally, we also determined the spectral dimension 
$d_s$ from the return probability of a random walker on the 
dual graph: $\rho(t) \sim t^{-d_s/2}$.
The time $t$ is measured in units of jumps between
neighboring simplexes, with a hopping probability $1/4$.
These estimates of $d_s$,
for volumes $N_3 = 500$, 2000, and 8000,
are included in Figure~\ref{fig.dh_pap}.
In the elongated phase we get $d_s \approx 4/3$, 
the value for a generic branched polymer, whereas   
in the crinkled phase the spectral dimension increases and 
reaches $d_s \approx 3/2$ for $\beta = -4$.  
Similar behavior is observed for the spectral
dimension in $4D$ \cite{us4d}, although in that case the estimate
in the crinkled phase was somewhat larger 
($d_s \approx 1.8$ for $\beta = -3.5$).
It is notable, however, that both in three and four dimensions
the spectral dimension in the crinkled phase is less than
two.

\section{Conclusions}

In this paper we have investigated 
an ensemble of maximal triangulations i.e.\ triangulations
that, for a given volume are as close to the upper kinematic
bound, $N_0/N_D = 1/D$, as possible. 
This is motivated by the observation 
that this ensemble appears to reflect
the properties of the weak-coupling limit of simplicial
gravity, $\kappa_0 \rightarrow \infty$.

In the first part of the paper, we presented an
algorithm which, in the unmodified case $\beta =0$,
yields an explicit enumeration of stacked spheres
--- a sub-class of the maximal ensemble.  We also 
generalized this algorithm to include a modified
measure, which, for $\beta \neq 0$, allows a recursive construction
of a SCE of the corresponding partition function 
Eq.~(\ref{part.measure}).  

Using a combination of SCE and MC simulations, we investigated 
the properties of the maximal ensemble.  We found that
it is composed out of three distinct series, all with
the same leading asymptotic behavior, but with
different leading corrections $\gamma(k)$.  This can be
understood as the maximal ensemble includes,
in addition to stacked spheres, triangulations that  are
``almost'' stacked spheres, 
i.e.\ stacked spheres with one or two defects.

In the second part we demonstrated that with a suitable modified
measure the maximal ensemble exhibits a transition to a crinkled
phase, akin to what is observed in $4D$ simplicial gravity.  This
agrees with the phase diagram Figure~1. The scaling properties of this
transition are consistent with a third order transition, i.e.\ with a
specific heat exponent $\alpha = -1$.  In the crinkled phase the
intrinsic fractal structure is charcterized by a string susceptibility
exponent $\gamma < 0$, and Hausdorff dimension $d_H = \infty$ or
$d_H^{\prime} \approx 2$, depending on whether it is measured on the
triangulations, or its dual.  This difference in the direct {\it vs}.\
dual fractal structure makes it unlikely that a thermodynamic limit
exist in the crinkled phase, where a theory of extended manifolds
emerges.  The value of te spectral dimension measured on the dual graph
agrees with the spectral dimension observed for the 
non--generic branched polymers. In the later case however 
exponent $\gamma$ is positive. In case of $\gamma<0$ BP model
predicts $d_s=2$ \cite{w}       

We note the similarity between the above phase structure 
and what is observed in two dimensions if one varies the measure.  
There too is a soft continuous transition to a crumpled phase, 
for sufficiently negative $\beta$, where the string susceptibility
exponent is negative \cite{2dcrump} and the internal 
geometry collapses.  Indeed, the similarity between $D=2$ and 
$D>2$ extends further.  In both cases, the same phase strucuture
is observed when sufficent amount of {\it matter} is 
coupled to the model.  This can be understood, qualitativly,
as the discretized form of the matter fields includes
to a leading order the measure Eq.~(\ref{part.measure}).

The phase diagram Figure~1 can also be partially understood from an
analysis of a mean-field model which is related to branched
polymers \cite{meanf}. Restricted to the maximal ensemble this
model predicts a soft (third order or greater) phase transition
between an elongated and a crinkled phase. In a language of 
branched polymers this corresponds to a transition from an elongated 
(or a generic) phase with $\gamma=1/2$ and $d_H=2$, to a bush phase 
(non-generic) with $\gamma=1-\beta$ and $d_H=\infty$. 
The intrinsic crumpling in the bush phase is though
much more pronounced than is observed in the crinkled phase 
of the maximal ensemble. 

To which extend do we expect the maximal ensemble to
correctly reflect the properties of the whole
weak-coupling phase, $\kappa_0 > \kappa_0^c$,
of the model Eq.~(\ref{part.canon}).
In principle, this may depend on how one
takes the two limits: the thermodynamic limit 
$N_3 \rightarrow \infty$, and the weak-coupling limit
$\kappa_0 \rightarrow \infty$.  A priori, these
two limits need not commute.  However, in 
Ref.~\cite{acm} it is claimed that already for
a finite $\kappa_0 > \kappa_0^c$, the partition function
Eq.~(\ref{part.canon}) is dominated by triangulations
which are {\it close} to stacked spheres.

\vspace{20pt}
\noindent
{\bf Acknowledgments:}
G.T.\  and P.B.\  were supported by the Alexander von Humboldt Foundation.

\newpage
\appendix

\section{Enumerations of the maximal ensemble}

In Tables~2 and 3 we list the first 20 terms in the SCE
of both labeled and unlabeled staked spheres in three and four
dimensions.
The enumeration of unlabeled stacked spheres
is given by the generating function Eq.~(\ref{unlab}), where:

{\small
\begin{eqnarray}
 \label{hh4}
 h_4(t) &= 
  & - \;{5\over 16} 
    + \;{1\over 3}  E_1\! \left(t\,E_4(t^3)\right ) 
    - \;{1\over 6}  E_1\! \left(t\,E_4(t^3)\right )^2
    + \;{1\over 4}  E_2\! \left(t\,E_4(t^2)\right )   \\ 
 && - \;{1\over 8}  E_2\! \left(t\,E_4(t^2)\right )^2
    + \;{1\over 6}  E_1\! \left(t^2\,E_4(t^6)\right )
                    E_2\! \left(t^3\,E_4(t^6)\right )  
    + \;{1\over 24}  E_4(t) 
      \nonumber \\ 
 && - \;{1\over 48}  E_4(t)^2
    + \;{1\over 120} t\,E_4(t)^5
    + \;{1\over 12}  t\,E_2\!\left (t\,E_4(t^2) \right )^3
                        E_4\!\left (t^2 \right ) 
    + \;{1\over 8}   t\,E_4\!\left (t^2 \right )^2
      \nonumber \\  
 && + \;{13\over 48} t^2\,E_4\!\left (t^2 \right )^4
    + \;{1\over 6}   t\,E_1\!\left (t\,E_4(t^3) \right )^2
                        E_4\!\left (t^3 \right ) 
    + \;{1\over 4}   t\,E_4\!\left (t^4 \right ) 
      \nonumber \\ 
 && + \;{1\over 4}   t^2\,E_4\!\left (t^4\right )^2
    + \;{1\over 5}   t\,E_4\!\left (t^5\right )
    + \;{1\over 6}   t^2\,E_1\!\left (t^2\,E_4(t^6) \right )
                        \,E_4\!\left (t^6 \right ) \,,
 \nonumber \\[20pt]
 h_5(t) &= 
   & - \;{19\over 60} 
     + \;{1\over 8}   E_1\! \bigl (t\,E_5(t^2)^2\bigr )
     - \;{1\over 16}  E_1\! \left (t\,E_5(t^2)^2\right )^2
     + \;{1\over 4}   E_1\! \left (t\,E_5(t^4)\right ) \\[2pt]
  && - \;{1\over 8}   E_1\! \left (t\,E_5(t^4)\right )^2
     + \;{1\over 6}   E_2\! \left (t\,E_5(t^3)\right )
     - \;{1\over 12}  E_2\! \left (t\,E_5(t^3)\right )^2
     + \;{1\over 12}  E_3\! \left (t\,E_5(t^2)\right )
      \nonumber \\[2pt] 
  && - \;{1\over 24}  E_3\! \left (t\,E_5(t^2)\right )^2
     + \;{1\over 6}  \;t\,E_2 \left (t^2\,E_5(t^6) \right )
                        \,E_3 \left (t^3\,E_5(t^6) \right )
     + \;{1\over 120} E_5(t)
      \nonumber \\[2pt]	      
  && + \;{1\over 12} \;t^2\,E_2 \left (t^2\,E_5(t^6) \right )^2
                          \,E_3 \left (t^3\,E_5(t^6) \right )^2
     - \;{1\over 240} E_5(t)^2
     + \;{1\over 720} \;t\,E_5(t)^6
      \nonumber \\[2pt]
  && + \;{1\over 48} \;t\,E_3 \left (t\,E_5(t^2) \right )^4
                        \,E_5 \bigl (t^2 \bigr ) 
     + \;{1\over 16} \;t\,E_1 \left (t\,E_5(t^2)^2 \right )^2
                       \; E_5 \bigl (t^2)^2
      \nonumber \\[2pt]
  && + \;{13\over 120} \,t^2\,E_5 \bigl (t^2 \bigr )^5
     + \;{1\over 18}  \,t\,E_2 \bigl (t\,E_5(t^3) \bigr )^3
                          \,E_5 \bigl (t^3 \bigr )
     + \;{1\over 18}  \,t\,E_5 \bigl (t^3 \bigr )^2
       \nonumber \\[2pt]
  && + \;{1\over 8} \;t\,E_1 \bigl (t\,E_5(t^4) \bigr )^2
                      \,E_5 \bigl (t^4 \bigr )
     + \;{1\over 8} \;t\,E_1 \bigl (t^2\,E_5(t^4)^2 \bigr )
                       \,E_5 \bigl (t^4 \bigr )
     + \;{1\over 48} \;t\,E_5 \bigl (t^2 \bigr )^3
       \nonumber \\[2pt] 
  && + \;{1\over 8} \;t^2\,E_1 \bigl (t^2\,E_5(t^4)^2 \bigr )
                         \,E_5 \bigl (t^4 \bigr )
     + \;{1\over 5}  \;t   \,E_5 \bigl (t^5 \bigr )
     + \;{1\over 10} \;t^2 \,E_5 \bigl (t^5 \bigr )^2
       \nonumber \\[2pt]
  && + \;{1\over 6} \;t    \,E_5 \bigl (t^6 \bigr )
     + \;{1\over 6} \;t^2 \,E_2 \bigl (t^2\,E_5(t^6) \bigr )^2
                           \,E_5 \bigl (t^6 \bigr )
     + \;{1\over 10} \;t^2 \,E_5 \bigl (t^{10} \bigr ) \,.
       \nonumber
\end{eqnarray}}

\setlength{\tabcolsep}{12pt}
\begin{table}
 \label{tab.enum3}
 \caption{\small The first 20 terms in the SCE of the
  both labeled and unlabeled three dimensional stacked spheres, 
  $S^0(N_3)$ and $h_{4,n}$, respectively.  
  This is for $\beta = 0$.}
 {\small
  \begin{center}
  \begin{tabular}{|r|c|r|r|} \hline
   \vspace{-8pt} & &  &   \\
   {\large $n$} & {\large $\bigl (N_0,N_3\bigr )$ }
     & \multicolumn{1}{c|}{\large $h_{4,n}$} 
     & \multicolumn{1}{c|}{\large$S^0(N_3)$} \\  
   \vspace{-8pt} & &  &  \\ \hline 
   \vspace{-8pt} & &  &  \\
    1 & (5,5) & 1 & 1\\
    2 & (6,8) & 1 & 5/2\\
    3 & (7,11) & 1 & 10\\
    4 & (8,14) & 3 & 50\\
    5 & (9,17) & 7 & 285\\
    6 & (10,20) & 30 & 1771\\
    7 & (11,23) & 131 & 11700\\
    8 & (12,26) & 795 & 80910\\
    9 & (13,29) & 5152 & 579700\\
   10 & (14,32) & 36800 & 8544965/2\\
   11 & (15,35) & 272093 & 32224114\\
   12 & (16,38) & 2077909 & 247754390\\
   13 & (17,41) & 16176607 & 1936016950\\
   14 & (18,44) & 127996683 & 15340200750\\
   15 & (19,47) & 1025727646 & 123020557800\\
   16 & (20,50) & 8310377720 & 996993749142\\
   17 & (21,53) & 67967600763 & 8155209690540\\
   18 & (22,56) & 560527576100 & 67259885983090\\
   19 & (23,59) & 4656993996246 & 558826638901000\\
   20 & (24,62) & 38949328897318 & 4673871155753800\\[4pt]\hline
  \end{tabular}
  \end{center}
 }
\end{table}

\setlength{\tabcolsep}{12pt}
\begin{table}
 \label{tab.enum4}
 \caption{\small Same as Table~2, except for four
  dimensional stacked spheres.}
 {\small
  \begin{center}
  \begin{tabular}{|r|c|r|r|} \hline
   \vspace{-8pt} & &  &   \\
   {\large $n$} & {\large $\bigl (N_0,N_4\bigr )$} 
       & \multicolumn{1}{c|}{\large $h_{5,n}$} 
       & \multicolumn{1}{c|}{\large $S^0(N_4)$} \\  
   \vspace{-8pt} & &  &  \\ \hline 
   \vspace{-8pt} & &  &  \\
    1 & (6,6) & 1 & 1\\
    2 & (7,10) & 1 & 3\\
    3 & (8,14) & 1 & 15\\
    4 & (9,18) & 3 & 95\\
    5 & (10,22) & 7 & 690\\
    6 & (11,26) & 30 & 5481\\
    7 & (12,30) & 142 & 46376\\
    8 & (13,34) & 922 & 411255\\
    9 & (14,38) & 6848 & 3781635\\
   10 & (15,42) & 57994 & 35791910\\
   11 & (16,46) & 525048 & 346821930\\
   12 & (17,50) & 4999697 & 3427001253\\
   13 & (18,54) & 49159506 & 34425730640\\
   14 & (19,58) & 494873071 & 350732771160\\
   15 & (20,62) & 5068890252 & 3617153918640\\
   16 & (21,66) & 52632367550 & 37703805776935\\
   17 & (22,70) & 552579655767 & 396716804816265\\
   18 & (23,74) & 5855580019967 & 4209161209968825\\
   19 & (24,78) & 62548009026369 & 44993046668984145\\
   20 & (25,82) & 672818970206219 & 484176486362971710\\[4pt]\hline
  \end{tabular}
  \end{center}
 }
\end{table}

\noindent
The values in Tables~2 and 3 can be compared to Table~8 of 
Ref.~\cite{hrs}. 
Although for most volumes they agree there are few discrepancies.
We have, however, an independent confirmation on our calculations;
for $N_D \leq 38$ we have explicitly identified 
all distinct triangulations.

\begin{figure}
 \epsfxsize=4in \centerline{ \epsfbox{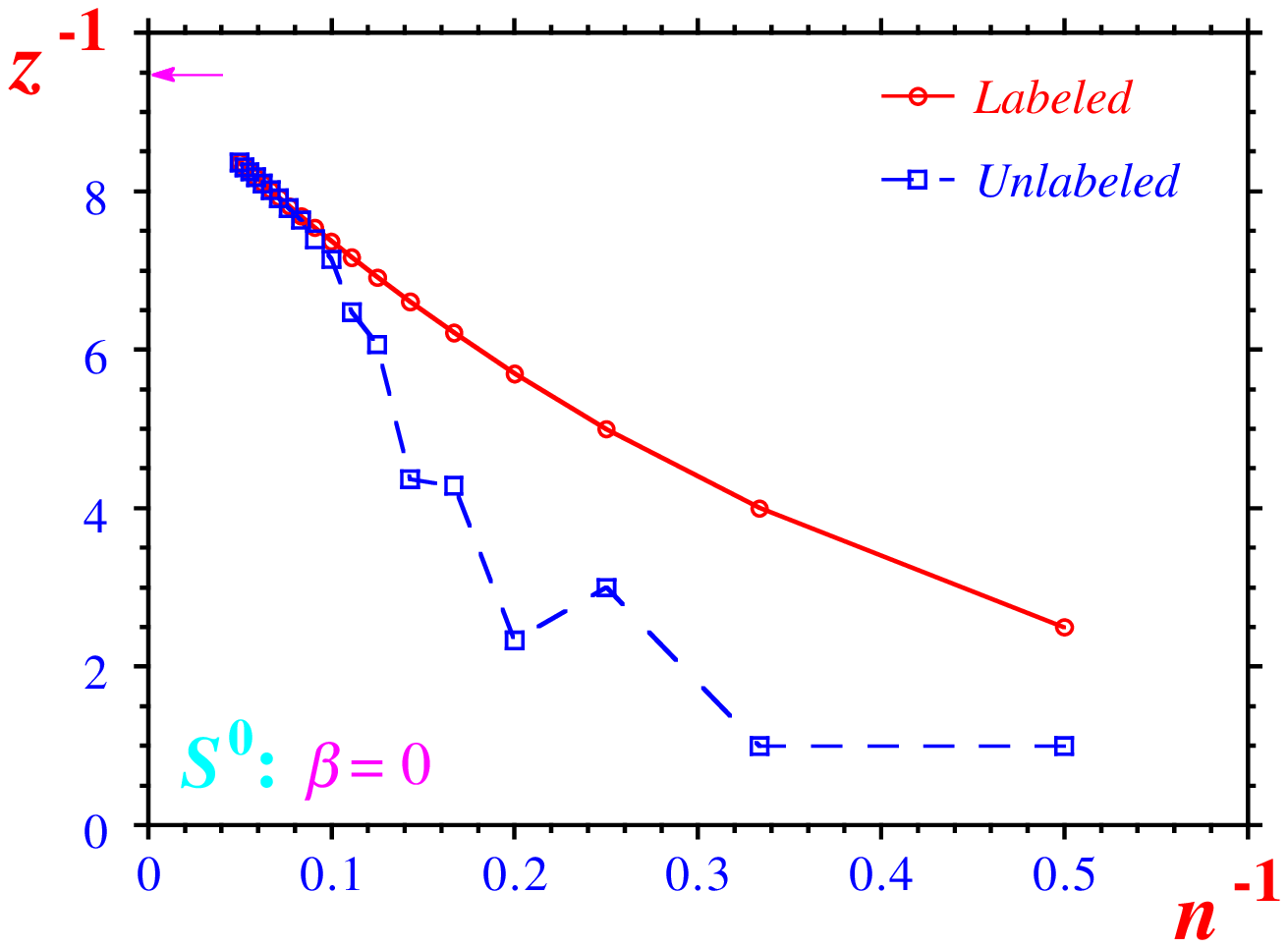}}
 \caption[fig.zeta_c]{\small The successibe ratios, 
  $s_n^{-1} = S_n/S_{n-1}$, for the sequence of labeled {\it vs}.\ 
  unlabeled stacked spheres. The arrow indicates the correct
  asymptotic value $z^{-1} = 256/27$.  For unlabeled triangulations 
  the irregular behavior is due to non-analytic corrections.}
 \label{fig.zeta_c}
\end{figure}

As discussed in Section~4.1, although the leading asymptotic
behavior is identical for the ensemble of labeled
{\it vs}.\ unlabeled stacked spheres, for the latter
the sub-leading corrections are non-analytic.  This effect,
which is demonstrated in Figure~(\ref{fig.zeta_c}),
makes it virtually impossible to extract the asymptotic
behavior of unlabeled triangulations using series extrapolation 
methods.

In Table~4 we list the first 10 and 8 terms of
two non-minimal series, $S^1$ and $S^2$, calculated
by directly identifying all distinct triangulations
in MC simulations.  

\setlength{\tabcolsep}{12pt}
\begin{table}
 \label{tab.SCEs1s2}
 \caption{\small The first few terms in the SCE
  of the $3D$ maximal ensemble for the two non-minimal 
  series $S^1$ and $S^2$.  $n_k$ is the number
  of distinct (unlabeled) triangulations.}
 {\small
  \begin{center}
  \begin{tabular}{|r|rrr|rrr|} \hline
   \vspace{-8pt} & & & & & &    \\
    $N_0$  & $N_3$ & $n_1$  & $S^1(N_3)$ 
           & $N_3$ & $n_2$  & $S^2(N_3)$ \\  
   \vspace{-8pt} & & & & & &  \\ \hline 
   \vspace{-8pt} & & & & & &  \\
     6  &   9  &      1  &      5/3  &  &  &  \\
     7  &  12  &      2  &     35/2
        &  13  &      1  &       15  \\
     8  &  15  &      5  &      152
        &  16  &      8  &  4205/16  \\
     9  &  18  &     23  &   3800/3 
        &  19  &     45  &     3290  \\
    10  &  21  &    124  &  73370/7 
        &  22  &    385  &  72535/2  \\
    11  &  24  &    859  &  348075/4
        &  25  &   3435  &  375750   \\
    12  &  27  &   6518  &  6544720/9
        &  28  &  32710  & 15043835/4 \\
    13  &  30  &  52761  &  6121632
        &  31  & 312601  &  36867985  \\
    14  &  33  & 438954  & 570959025/11
        &  34  &2995589  &  713253255/2  \\
    15  &  36  &3717370  & 886163135/2
        &    &   &    \\[4pt]\hline
  \end{tabular}
  \end{center}
 }
\end{table}


\section{The contact term}

As discussed in Section~5.2, for minbu's on maximal
triangulations, 
the interactions at neck --- the {\it contact} term ---
play a (unusually) big and non-trivial role in the crinkled phase.  
It is essential to take these interactions into account when
estimating $\gamma$ from a fit to the measured
minbu distributions.  In this appendix we analyze the
behavior of the contact term and explain why it plays
such an important role for this particular ensemble.

In the nomenclature of Section~3 we can write the minbu
distribution, Eq.~(\ref{babymax}), for $\beta =0$ and 
measured on a stacked sphere of size $n$, as
\begin{equation}
 \nonumber
 {\cal B}_n(b) \;\propto \;\frac{e_{4,b}\,e_{4,n-b}}{e_{4, n}}  
 \;\propto\; \bigl (3(n\!-\!b) + 2 \bigr ) \bigl (3 b + 2 \bigr)\;
  \frac{W_3(b)W_3(n-b)}{W_3(n)}
\end{equation}
However, for $\beta\ne 0$ the situation is different. 
We must take into account the different possible 
distributions of vertex orders, ${\bf p}$ and ${\bf q}$,
on the marked simplex (the neck), both on the baby and on the mother. 
The distribution is then 
\begin{equation}
 \label{bab}
 \widetilde{\cal B}_n(b;\beta) \;\propto\; \frac{1}{W_3(n)}\;
 \sum_{{\bf p},{\bf q}} \;\;\prod_{i=1}^4 \;(p_i+q_i-2)^\beta
 \;e({\bf p};n\!-\!b,\beta)\; e({\bf q};b,\beta)\;.
\end{equation}
Defining 
\begin{equation}	
 w(\mathbf{q};n,\beta) \;=\; 
 \frac{e(\mathbf{q};n,\beta)}{e(n,\beta)}
 \,\prod_{i=1}^4 q_i^\beta \;,
\end{equation}
we can rewrite Eq.~(\ref{bab}) as
\begin{eqnarray}
\label{babw}
 \label{ct} \nonumber
 \widetilde{\cal B}_n(b;\beta)
  &\propto &\bigl (3(n-b) + 2 \bigr ) \bigl (3 b + 2 \bigr)\,
   \frac{W_3(b)\;W_3(n\!-\!b)}{W_3(n)} \\ \nonumber
  && \;\;\;\;\times \sum_{{\bf q},{\bf p}}
   \;w({\bf q};n\!-\!b,\beta)\,w({\bf p};b,\beta)
   \,\prod_{i=1}^4 \frac{(q_i+p_i-2)^\beta}{q_i^\beta p_i^\beta} \\[4pt]
   &\equiv  &{\cal B}_n(b;\beta)\cdot {\cal C}_n(b;\beta)\;,
\end{eqnarray}
where we define ${\cal C}_n(b;\beta)$ as the {\it contact} 
term.. 
 
The following argument gives some idea of the relevance 
of the contact term in the different phases.
If we assume that the distribution $w({\bf q})$ factorizes,
\begin{equation}
 w({\bf q};n,\beta) \;=\; \prod_{i=1}^4 \, w(q_i;n,\beta)\;,
\end{equation}
then the distribution $w(q;n)$ is related to the vertex order 
distribution $\pi(q)$, i.e.\ the probability of finding a vertex 
with given order $q$, through
\begin{equation}
 w(q;n,\beta) \;=\; \frac{q \, \pi(q;n,\beta)}{\langle q \rangle}.
\end{equation}
In the elongated phase the distribution $w(q;n)$ 
decays exponentially, whereas in the crinkled phase 
vertices of large order are favored and the distribution 
has a power-law decay (Figure~\ref{fig.vord}). 

\begin{figure}
 \begin{center}
  \mbox{\psfig{figure=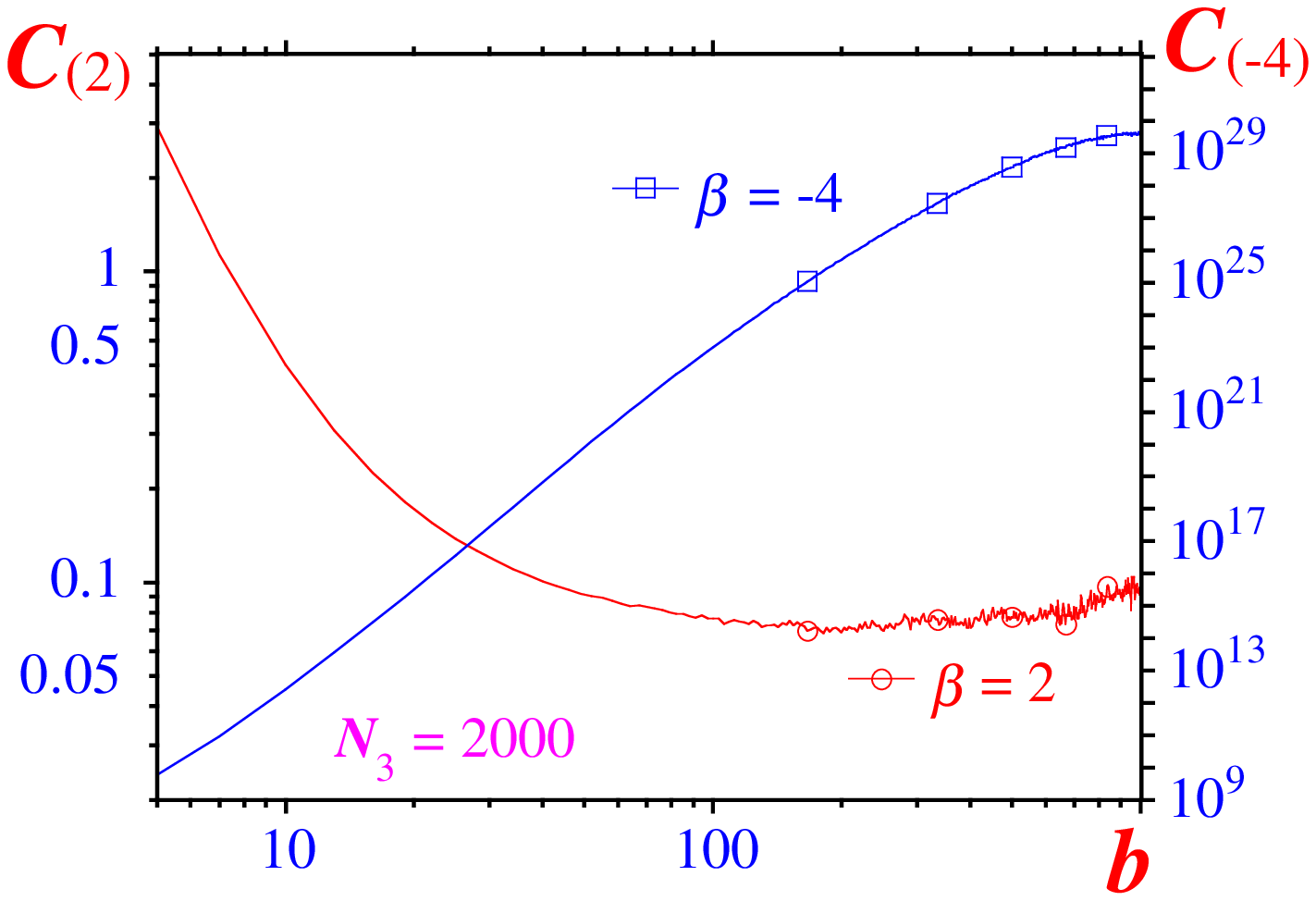,width=6.80cm}  
        \psfig{figure=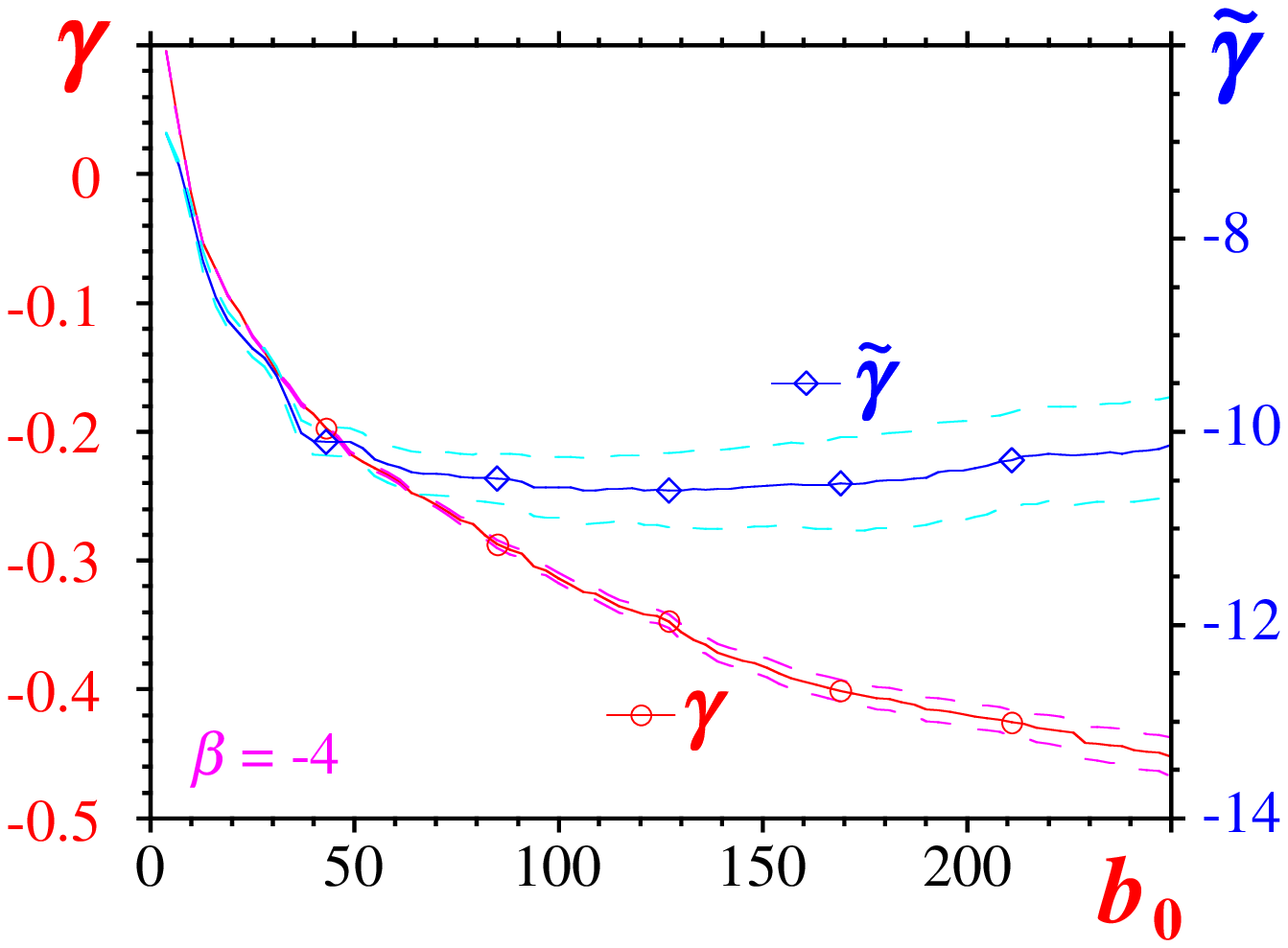,width=6.80cm} }     
 \end{center}
 \caption{\small ({\sf Left}) The dependence of the contact term,
  ${\cal C}_n(b;\beta)$, on the minbu size $b$.
  This is for $N_3 = 2000$ and both in the elongated phase, $\beta = 2$, 
  and in the crinkled phase, $\beta = -4$: $C(2)$ and $C(-4)$. 
  ({\sf Right}) The exponent $\gamma$, determined by a fit to 
  the minbu distribution Eq.~(\ref{babymax})
  for $N_3 = 2000$ and $\beta = -4$. 
  Results are shown varying a lower cut-off $b_0$
  in the fit, and both with and without the contact term
  divided out, $\widetilde{\gamma}$ and $\gamma$ respectively.
  Note that in both plots the compared curves are plotted on
  very different scales.}
 \label{fig.cont}
\end{figure}

An exponential decay of $w(q;n)$ effectively
introduces a cut-off in the sum over vertex orders in Eq.~(\ref{ct}).
Hence, for sufficiently large triangulations the contact 
term becomes independent of the volume. 
In contrast, in the crinkled phase  
there is no such cut-off and we can expect 
a strong volume dependence.  This is demonstrated by numerical
measurements of the contact term, shown in Figure~\ref{fig.cont}.
In the elongated phase, $\beta = 2$, the contact term ${\cal C}_n(2)$
is more or less constant for $\beta \gtrsim 50$,  whereas for
$\beta = -4$ the contact term $C_n(-4)$ depends strongly on $b$.

Although this effect can be expected for any ensemble 
modified by the measure Eq.~(\ref{part.measure}),
the particular construction of maximal triangulations,
by a repeated application of move $\left(1,4\right)$, 
enhances the role of the contact term.
For every time move $\left(1,4\right)$
is applied one minimal neck is generated, hence for
a maximal triangulation the number of minbu's is fixed.
Thus the minbu distributions, measured on $S^0$,
is more constrained than measured on
a general ensemble where triangulations with
fewer minbu's are favored as $\beta \rightarrow \infty$.

How does all this affect the measurements of $\gamma\;$?
The standard method for extracting $\gamma$ from a 
minbu distribution is to impose a lower
cut-off $b_0$ on the minbu size included in the fit to
Eq.~(\ref{babymax}).  The cut-off is then increased until a stable
estimate of $\gamma(b_0)$ is obtained \cite{minbu2}.
In the elongated phase, and in general for most models
of simplicial gravity, this methods works fine.
However, for stacked spheres this method fails in the crinkled
phase.  Unless the contact term is eliminated, i.e.\
divided out of the minbu distribution in the measurement
process, a stable plateau in $\gamma(b_0)$ is not observed.  
An example of this is shown in Figure~\ref{fig.cont} 
for $\beta = -4$.

\end{document}